\pdfoutput=1
\newif\ifplainstyle
\plainstyletrue
\plainstylefalse

\newif\ifjhepstyle
\jhepstyletrue

\newif\ifprstyle
\prstyletrue
\prstylefalse

\ifprstyle
	\documentclass[twocolumn,nofootinbib]{revtex4-1}
\else
	\documentclass[11pt,a4paper]{article}
\fi

\ifjhepstyle
	\usepackage{jheppub}
	\usepackage{amsfonts}
	\usepackage{verbatim}
	\usepackage{float}
	\usepackage{color}
	\usepackage{array}
	\newcolumntype{C}[1]{>{\centering\arraybackslash$}p{#1}<{$}}
	\makeatletter
	\def\@fpheader{\phantom{:-)}}
	\makeatother
\else	
 	\ifprstyle
		\usepackage{verbatim}
		\usepackage{amsmath,amsfonts,amssymb}
		\usepackage[colorlinks=true
                	,urlcolor=blue
                	,anchorcolor=blue
                	,citecolor=blue
                	,filecolor=blue
                	,linkcolor=blue
                	,menucolor=blue
                	]{hyperref}
	\else
            	\usepackage{verbatim}
            	\usepackage{cite}
            	\usepackage{setspace}
            	\usepackage[top=2.5cm, bottom=2.75cm, left=2.5cm, right=2.5cm]{geometry}
            	\usepackage{amsmath,amsfonts,amssymb}
            	\usepackage[colorlinks=true
            	,urlcolor=blue
            	,anchorcolor=blue
            	,citecolor=blue
            	,filecolor=blue
            	,linkcolor=blue
            	,menucolor=blue
            	,linktoc=page
            	]{hyperref}
            	\usepackage{float}
            	\restylefloat{table}
            	\renewcommand{\arraystretch}{1.5}
            	\numberwithin{equation}{section}
	\fi
\fi

\usepackage{subfig}

\usepackage{etoolbox}

\makeatletter
\def\hlinewd#1{%
\noalign{\ifnum0=`}\fi\hrule \@height #1 %
\futurelet\reserved@a\@xhline}
\makeatother

\newcolumntype{?}[1]{!{\vrule width #1}}

\allowdisplaybreaks

\usepackage{multirow}

\usepackage[normalem]{ulem}
\usepackage{mathtools}
\usepackage{transparent}
\usepackage{bm}
\usepackage{enumitem}

\usepackage[textsize=scriptsize,textwidth=2.5cm]{todonotes}

\makeatletter
\let\save@mathaccent\mathaccent
\newcommand*\if@single[3]{%
  \setbox0\hbox{${\mathaccent"0362{#1}}^H$}%
  \setbox2\hbox{${\mathaccent"0362{\kern0pt#1}}^H$}%
  \ifdim\ht0=\ht2 #3\else #2\fi
  }
\newcommand*\rel@kern[1]{\kern#1\dimexpr\macc@kerna}
\newcommand*\widebar[1]{\@ifnextchar^{{\wide@bar{#1}{0}}}{\wide@bar{#1}{1}}}
\newcommand*\wide@bar[2]{\if@single{#1}{\wide@bar@{#1}{#2}{1}}{\wide@bar@{#1}{#2}{2}}}
\newcommand*\wide@bar@[3]{%
  \begingroup
  \def\mathaccent##1##2{%
    \let\mathaccent\save@mathaccent
    \if#32 \let\macc@nucleus\first@char \fi
    \setbox\z@\hbox{$\macc@style{\macc@nucleus}_{}$}%
    \setbox\tw@\hbox{$\macc@style{\macc@nucleus}{}_{}$}%
    \dimen@\wd\tw@
    \advance\dimen@-\wd\z@
    \divide\dimen@ 3
    \@tempdima\wd\tw@
    \advance\@tempdima-\scriptspace
    \divide\@tempdima 10
    \advance\dimen@-\@tempdima
    \ifdim\dimen@>\z@ \dimen@0pt\fi
    \rel@kern{0.6}\kern-\dimen@
    \if#31
      \overline{\rel@kern{-0.6}\kern\dimen@\macc@nucleus\rel@kern{0.4}\kern\dimen@}%
      \advance\dimen@0.4\dimexpr\macc@kerna
      \let\final@kern#2%
      \ifdim\dimen@<\z@ \let\final@kern1\fi
      \if\final@kern1 \kern-\dimen@\fi
    \else
      \overline{\rel@kern{-0.6}\kern\dimen@#1}%
    \fi
  }%
  \macc@depth\@ne
  \let\math@bgroup\@empty \let\math@egroup\macc@set@skewchar
  \mathsurround\z@ \frozen@everymath{\mathgroup\macc@group\relax}%
  \macc@set@skewchar\relax
  \let\mathaccentV\macc@nested@a
  \if#31
    \macc@nested@a\relax111{#1}%
  \else
    \def\gobble@till@marker##1\endmarker{}%
    \futurelet\first@char\gobble@till@marker#1\endmarker
    \ifcat\noexpand\first@char A\else
      \def\first@char{}%
    \fi
    \macc@nested@a\relax111{\first@char}%
  \fi
  \endgroup
}
\makeatother
\newcommand{\ThisIsTheTitle}{On the universal behavior of $T\bar T$-deformed CFTs:\\ single and double-trace partition functions at large $c$
} 
\newcommand{\ThisIsAuthorOne}{Luis Apolo,}
\newcommand{\ThisIsEmailOne}{l.a.apolo@uva.nl}
\newcommand{\ThisIsAuthorTwo}{Wei Song}
\newcommand{\ThisIsEmailTwo}{wsong2014@mail.tsinghua.edu.cn}
\newcommand{\ThisIsAuthorThree}{and Boyang Yu}
\newcommand{\ThisIsEmailThree}{yuby21@pku.edu.cn}
\newcommand{\TheseAreTheKeywords}{}
\newcommand{\ThisIsTheAbstract}{
We study universal properties of the torus partition function of  $T\bar T$-deformed CFTs  under the assumption of modular invariance, for both the original version, referred to as the double-trace version in this paper, and the single-trace version defined as the symmetric product orbifold of double-trace $T\bar T$-deformed CFTs. In the double-trace case, we specify sparseness conditions for the light states for which the partition function at low temperatures is dominated by the vacuum when the central charge of the undeformed CFT is large. Using modular invariance, this implies a universal density of high energy states, in analogy with the behavior of holographic CFTs. For the single-trace $T\bar T$ deformation, we show that modular invariance implies that the torus partition function can be written in terms of the untwisted partition function and its modular images, the latter of which can be obtained from the action of a generalized Hecke operator.  The partition function and the energy of twisted states  match holographic calculations in previous literature, thus providing further evidence for the conjectured holographic correspondence. In addition, we show that the single-trace partition function is universal when the central charge of the undeformed CFT is large, without needing to assume a sparse density of light states. Instead, the density of light states is shown to always saturate the sparseness condition.
}

\ifjhepstyle
\title{\ThisIsTheTitle}

\author[ a, b ]{\ThisIsAuthorOne}
\author[ c ]{\ThisIsAuthorTwo}
\author[ d ]{\ThisIsAuthorThree}

\affiliation[a]{Institute for Theoretical Physics, University of Amsterdam, 1090GL Amsterdam, The Netherlands}
\affiliation[b]{Beijing Institute of Mathematical Sciences and Applications, Beijing 101408, China}

\affiliation[c]{Yau Mathematical Sciences Center, Tsinghua University, Beijing 100084, China}

\affiliation[d]{Center for High Energy Physics, Peking University, Beijing 100871, China}

\emailAdd{\ThisIsEmailOne}
\emailAdd{\ThisIsEmailTwo}
\emailAdd{\ThisIsEmailThree}

\abstract{\ThisIsTheAbstract} 

\keywords{\TheseAreTheKeywords}
\fi

\begin{document}

\ifjhepstyle
\maketitle
\flushbottom
\fi

\long\def\symfootnote[#1]#2{\begingroup%
\def\thefootnote{\fnsymbol{footnote}}\footnote[#1]{#2}\endgroup} 

\def\rednote#1{{\color{red} #1}}
\def\bluenote#1{{\color{blue} #1}}
\def\magnote#1{{\color{magenta} #1}}
\def\tealnote#1{{\color{teal} #1}}
\def\rout#1{{\color{red} \sout{#1}}}

\newcommand{\wei}[2]{\textcolor{magenta}{#1}\todo[color=green]{\scriptsize{W: #2}}}
\newcommand{\lui}[2]{\textcolor{blue}{#1}\todo[color=yellow]{\scriptsize{L: #2}}}
\newcommand{\by}[2]{\textcolor{red}{#1}\todo[color=yellow]{\scriptsize{BY: #2}}}
\newcommand{\LA}[1]{{\color{blue}{\bf LA:} #1}}
\newcommand{\BY}[1]{{\bf \color{red}{\bf BY:} #1}}

\def\({\left (}
\def\){\right )}
\def\lb{\left [}
\def\rb{\right ]}
\def\lB{\left \{}
\def\rB{\right \}}

\def\Int#1#2{\int \textrm{d}^{#1} x \sqrt{|#2|}}
\def\Bra#1{\left\langle#1\right|} 
\def\Ket#1{\left|#1\right\rangle}
\def\BraKet#1#2{\left\langle#1|#2\right\rangle} 
\def\Vev#1{\left\langle#1\right\rangle}
\def\Vevm#1{\left\langle \Phi |#1| \Phi \right\rangle}\def\bbox{\bar{\Box}}
\def\til#1{\tilde{#1}}
\def\wtil#1{\widetilde{#1}}
\def\ph#1{\phantom{#1}}

\def\ra{\rightarrow}
\def\la{\leftarrow}
\def\lra{\leftrightarrow}
\def\p{\partial}
\def\barp{\bar{\partial}}
\def\diff{\mathrm{d}}

\def\sinh{\mathrm{sinh}}
\def\cosh{\mathrm{cosh}}
\def\tanh{\mathrm{tanh}}
\def\coth{\mathrm{coth}}
\def\sech{\mathrm{sech}}
\def\csch{\mathrm{csch}}

\def\a{\alpha}
\def\b{\beta}
\def\g{\gamma}
\def\d{\delta}
\def\e{\epsilon}
\def\ve{\varepsilon}
\def\k{\kappa}
\def\l{\lambda}
\def\n{\nabla}
\def\om{\omega}
\def\s{\sigma}
\def\t{\theta}
\def\z{\zeta}
\def\vp{\varphi}

\def\ss{\Sigma}
\def\dd{\Delta}
\def\GG{\Gamma}
\def\LL{\Lambda}
\def\tt{\Theta}

\def\A{{\mathcal A}}
\def\B{{\mathcal B}}
\def\C{{\mathcal C}}
\def\cE{{\mathcal E}}
\def\D{{\mathcal D}}
\def\F{{\mathcal F}}
\def\H{{\mathcal H}}
\def\I{{\mathcal I}}
\def\J{{\mathcal J}}
\def\K{{\mathcal K}}
\def\L{{\mathcal L}}
\def\M{{\mathcal M}}
\def\N{{\mathcal N}}
\def\O{{\mathcal O}}
\def\Q{{\mathcal Q}}
\def\P{{\mathcal P}}
\def\cS{{\mathcal S}}
\def\T{{\mathcal T}}
\def\W{{\mathcal W}}
\def\X{{\mathcal X}}
\def\Z{{\mathcal Z}}

\def\mfa{\mathfrak{a}}
\def\mfb{\mathfrak{b}}
\def\mfc{\mathfrak{c}}
\def\mfd{\mathfrak{d}}

\def\we{\wedge}
\def\re{\textrm{Re}}

\def\tilw{\tilde{w}}
\def\tile{\tilde{e}}

\def\tilL{\tilde{L}}
\def\tilJ{\tilde{J}}

\def\zz{\bar z}
\def\xx{\bar x}
\def\yy{\bar y}
\def\xp{x^{+}}
\def\xm{x^{-}}

\def\bp{\bar{\p}}
\def\note#1{{\color{red}#1}}
\def\notebf#1{{\bf\color{red}#1}}

\def\vac{\text{vac}}

\def\VirU1{Vir \times U(1)}
\def\VirSL2R{\mathrm{Vir}\otimes\widehat{\mathrm{SL}}(2,\mathbb{R})}
\def\U1{U(1)}
\def\u1{U(1)}
\def\SL2R{\widehat{\mathrm{SL}}(2,\mathbb{R})}
\def\sl2r{\mathrm{SL}(2,\mathbb{R})}

\def\RR{\mathbb{R}}

\def\tr{\mathrm{Tr}}
\def\bnabla{\overline{\nabla}}

\def\sint{\int_{\ss}}
\def\dsint{\int_{\p\ss}}
\def\hint{\int_{H}}

\def\sym{\textrm{Sym}}
\def\symTT{\textrm{Sym}^N \mathcal M_\mu}
\def\symCFT{\textrm{Sym}^N \mathcal M_0}
\def\seedTT{\mathcal M_\mu}
\def\seedCFT{\mathcal M_0}

\newcommand{\eq}[1]{\begin{align}#1\end{align}}
\newcommand{\eqst}[1]{\begin{align*}#1\end{align*}}
\newcommand{\eqsp}[1]{\begin{equation}\begin{split}#1\end{split}\end{equation}}

\newcommand{\absq}[1]{{\textstyle\sqrt{\left |#1\right |}}}



\newcommand{\pp}{{\partial_+}{}}
\newcommand{\pmm}{{\partial_-}{}}

\newcommand{\be}{\begin{equation}}
\newcommand{\ee}{\end{equation}}
\newcommand{\bea}{\begin{eqnarray}}
\newcommand{\eea}{\end{eqnarray}}
\newcommand{\bb}{\mathbb}
\newcommand{\ba}{\begin{aligned}}
\newcommand{\ea}{\end{aligned}}
\newcommand{\me}{\mathcal{E}}


\ifprstyle
\title{\ThisIsTheTitle}

\author{\ThisIsAuthorOne}
\email{\ThisIsEmailOne}

\author{\ThisIsAuthorTwo}
\email{\ThisIsEmailTwo}

\affiliation{\ThisIsTheAffiliation}


\begin{abstract}
\ThisIsTheAbstract
\end{abstract}


\maketitle

\fi
\ifplainstyle
\begin{titlepage}
\begin{center}

\ph{.}

\vskip 4 cm

{\Large \bf \ThisIsTheTitle}

\vskip 1 cm

\renewcommand*{\thefootnote}{\fnsymbol{footnote}}

{{\ThisIsAuthorOne}\footnote{\ThisIsEmailOne} } and {\ThisIsAuthorTwo}\footnote{\ThisIsEmailTwo}

\renewcommand*{\thefootnote}{\arabic{footnote}}

\setcounter{footnote}{0}

\vskip .75 cm


\end{center}

\vskip 1.25 cm
\date{}


\end{titlepage}

\newpage

\fi

\ifplainstyle
\tableofcontents
\noindent\hrulefill
\bigskip
\fi

\section{Introduction and summary} \label{se:introduction}

The seminal work \cite{Strominger:1996sh} took an important step in unravelling the mystery of black hole physics by providing a microscopic account of the Bekenstein-Hawking entropy. The essence of \cite{Strominger:1996sh} lies in the fact that the Bekenstein-Hawking entropy of BTZ black holes agrees with Cardy's formula for the asymptotic density of states in two-dimensional CFTs \cite{Strominger:1997eq}, which is now a standard entry in the dictionary of the AdS/CFT correspondence. One subtlety of the aforementioned matching is that holographic CFTs feature a large central charge $c\gg1$ and semiclassical black holes correspond to states with energies comparable to, but not necessarily much larger than $c$. In contrast, Cardy's formula works for states with energies $E \gg c$ and fixed central charge \cite{Cardy:1986ie}. Nevertheless, the range of validity of Cardy's formula can be extended to the holographic regime where $c \gg 1$ by assuming that the density of light states is sparse  \cite{Hartman:2014oaa}. The study of holographic CFTs featuring a large central charge and a sparse spectrum of light states, among other properties, has deepened our understanding of holographic dualities and the space of conformal field theories.

While the microscopic counting of \cite{Strominger:1996sh} can be understood within the AdS$_3$/CFT$_2$ correspondence, many interesting black holes, including the ones found in the real world, are not asymptotically AdS$_3$.\footnote{See \cite{Benini:2015eyy,Cabo-Bizet:2018ehj,Choi:2018hmj,Benini:2018ywd} for recent developments in higher dimensions and \cite{Guica:2008mu} for Kerr black holes.} On the gravity side, the Bekenstein-Hawking entropy of these black holes depends only on gravity in the near-horizon geometry and is not sensitive to the region of spacetime that is far away from the horizon. This suggests that some universal properties of the dual CFT might be preserved under irrelevant deformations, the latter of which modify the asymptotic region of the spacetime. In general, it is difficult to keep track of deformations driven by irrelevant operators. Nevertheless, in two dimensions, a family of solvable irrelevant deformations exists that is driven by the antisymmetric product of conserved currents \cite{Smirnov:2016lqw}, and includes the $T\bar T$ \cite{Zamolodchikov:2004ce,Smirnov:2016lqw,Cavaglia:2016oda} and $J\bar T$ \cite{Guica:2017lia} deformations, as well as linear combinations of them \cite{LeFloch:2019rut,Chakraborty:2019mdf,Frolov:2019xzi}. These models are interesting from field theoretical and holographic points of view, and have been studied extensively in recent years. In this paper, we explore universal features of $T\bar T$-deformed CFTs with a particular focus on the holographic regime. 

The $T\bar T$ deformation describes a one-parameter family of quantum field theories via the differential equation \cite{Zamolodchikov:2004ce,Smirnov:2016lqw,Cavaglia:2016oda}
\eq{
\frac{\p I}{\p\mu} = 8\pi \int d^2 x \, T\bar T, \qquad T\bar T \coloneqq \frac{1}{8} \big( T^{\a\b} T_{\a\b} - (T_{\a}^\a)^2 \big), \label{TTbardefinition}
}
where $I$ and $T_{\a\b}$ are the action and the stress tensor of the deformed theory at deformation parameter $\mu$. Despite being irrelevant,  the $T\bar T$ deformation is still solvable in the sense that spectrum on the cylinder can be expressed in terms of the undeformed spectrum and the deformation parameter~\cite{Zamolodchikov:2004ce,Smirnov:2016lqw,Cavaglia:2016oda}, and moreover the deformed S-matrix can be obtained by dressing the undeformed one with a CDD factor \cite{Dubovsky:2012wk,Dubovsky:2013ira,Dubovsky:2017cnj}. Another remarkable feature of $T\bar T$-deformed CFTs is that the torus partition function remains modular invariant \cite{Datta:2018thy,Aharony:2018bad}, despite the fact that conformal  symmetry is broken by the deformation. Other interesting features of $T\bar T$-deformed QFTs include connections to two-dimensional gravity and string theory~\cite{Dubovsky:2012wk,Cavaglia:2016oda,Dubovsky:2017cnj,Cardy:2018sdv,Callebaut:2019omt,Tolley:2019nmm}, the computation of correlation functions and entanglement entropy~\cite{Giribet:2017imm,Donnelly:2018bef,Chen:2018eqk,Jeong:2019ylz,Cardy:2019qao}, and generalizations to higher and lower dimensions~\cite{Hartman:2018tkw,Taylor:2018xcy,Gross:2019ach}, among others. 

By definition, the $T\bar T$ deformation \eqref{TTbardefinition} is a double-trace deformation that can be applied to any quantum field theory with a well-defined stress tensor~\cite{Zamolodchikov:2004ce}. A single-trace version of the deformation can be defined for theories obtained from the product of QFTs, and in particular, to symmetric product orbifold CFTs \cite{Giveon:2017nie}. A symmetric product orbifold CFT, denoted by $\symCFT \coloneqq (\seedCFT)^N/S_N$, consists of $N$ copies of a seed CFT $\seedCFT$ supplemented by the condition that all states are invariant under the symmetric group $S_N$. The central charge of this theory is $c = N c_0$ where $c_0$ denotes the central charge of the seed CFT $\seedCFT$. The single-trace $T\bar T$ deformation of $\symCFT$ yields another symmetric product orbifold, $\symTT$, whose seed theory $\seedTT$ is the $T\bar T$ deformation of the seed CFT $\seedCFT$. In other words, under the single-trace deformation each copy of the seed CFT is deformed by the $T\bar T$ operator in that copy,
\eq{ 
\frac{\p I^{(i)}}{\p\mu} = 8\pi \int d^2 x \,  \, (T\bar T)^{(i)}, \qquad (T\bar T)^{(i)} \coloneqq \frac{1}{8} \big( T^{(i)\a\b} T^{(i)}_{\a\b} - (T_{\a}^{(i)\a})^2 \big), \label{TTbardefinition2} 
} 
where $I^{(i)}$ is the action of the $i$th copy of $\symTT$. In order to distinguish the original $T\bar T$ deformation \eqref{TTbardefinition} from the single-trace version \eqref{TTbardefinition2}, we sometimes refer to the former as the double-trace $T\bar T$ deformation. 

Double-trace $T\bar T$-deformed CFTs have been argued to be dual to semiclassical Einstein gravity with a negative cosmological constant, with the metric satisfying Dirichlet boundary conditions on a cutoff surface \cite{McGough:2016lol} or, equivalently, mixed boundary conditions at the asymptotic boundary \cite{Guica:2019nzm}. On the other hand, single-trace $T\bar T$-deformed CFTs have been argued to be dual to the long string sector of string theory on three-dimensional linear dilaton \cite{Giveon:2017nie} and TsT-transformed backgrounds \cite{Apolo:2019zai} (see also~\cite{Apolo:2018qpq,Araujo:2018rho,Borsato:2018spz,Apolo:2021wcn} for the relation between TsT transformations and more general single-trace irrelevant deformations).

In this paper, we study universal features of double and single-trace $T\bar T$-deformed CFTs that follow from modular invariance of the torus partition function. Throughout this paper, we will assume that the $T\bar T$ deformation parameter is positive, which in our conventions means that the deformed energies are always real. In particular, we will explore the regime where the central charge of the undeformed CFTs is large, which is a necessary but not sufficient condition for a holographic description in terms of a semiclassical theory of gravity. The main results of this paper are summarized as follows.

\bigskip

\noindent {\bf Double-trace $T\bar T$-deformed CFTs.} In analogy with the analysis of Hartman, Keller, and Stoica (HKS) in two-dimensional CFTs \cite{Hartman:2014oaa}, we will show that the partition function $Z(\tau, \bar\tau; \mu)$ of double-trace $T\bar T$-deformed CFTs is universal at large $c$ provided that the spectrum of light states is sparse in a sense that we shall make precise. In this case, the partition function  is found to satisfy
\eq{
\log Z(\tau, \bar\tau; \mu) \approx \max \bigg\{\pi i (\tau - \bar\tau) E_{\vac}(\mu), -\pi i \Big (\frac{1}{\tau} - \frac{1}{\bar\tau}\Big) E_{\vac}\Big(\frac{\mu}{\tau\bar\tau}\Big)\bigg\}, \qquad |\tau|^2 \ne 1, \label{intro1}
}
where $\tau$ is the modular parameter and $E_\vac(\mu)$ is the energy of the vacuum. Eq.~\eqref{intro1} tells us that at temperatures $|\tau|^2 > 1$, the partition function is dominated by the vacuum, while at temperatures $|\tau|^2 <1$, it is given by its modular image ($\tau \mapsto -1/\tau$). The analog of the Hawking-Page phase transition takes place at $|\tau |^2 = 1$, where the bulk saddles corresponding to thermal AdS$_3$ and the BTZ black hole exchange dominance. Transforming \eqref{intro1} to the microcanonical ensemble, the logarithm of the asymptotic density of states is given by the $T\bar T$ analog of the Cardy formula
\eq{
S(E_L, E_R)  & \approx 2\pi \bigg \{ \sqrt{\frac{c}{6} E_L(\mu) \Big(1 +  {2\mu} E_R(\mu) \Big)} + \sqrt{\frac{c}{6} E_R(\mu) \Big( 1 + {2\mu} E_L(\mu) \Big) }\, \bigg \}, \label{entropy}
}
where $c$ is the central charge of the undeformed CFT while the left and right-moving energies $E_L$ and $E_R$ are given in terms of the energy $E$ and the angular momentum $J$ by $E_L = \frac{1}{2}(E + J)$ and $E_R = \frac{1}{2}(E - J)$. The entropy \eqref{entropy} is valid when
 \eq{
\frac{E_L(\mu)}{c} \frac{E_R(\mu)}{c} > \frac{1}{24^2}\big(1 + \mathcal O(c \mu) \big), \qquad c\gg1, \label{holographic}
}
where $c\mu$ is held fixed in the limit where $c$ is large, as described in more detail later.

The entropy \eqref{entropy} can be alternatively obtained by rewriting the Cardy formula for the undeformed CFT in terms of the energies of the $T\bar T$-deformed theory. This can be justified by the observation that the energy levels do not cross each other under the $T\bar T$ deformation \cite{McGough:2016lol}. The result \eqref{entropy} was also derived in \cite{Datta:2018thy} for states in the Cardy regime where $E_{L,R}(\mu) \gg c$ as in \cite{Cardy:1986ie}; hence, it applies holographically to very large black holes, whose energies are much greater than $c$. In contrast, the range of validity of \eqref{holographic} is more apt for comparison with holography, since a large central charge in the undeformed CFT translates to a small gravitational coupling in the bulk, where the Bekenstein-Hawking computation is reliable. In this case, the partition function \eqref{intro1} is expected to reproduce the free energy of black holes with energies greater than $c$, but not necessarily much larger than $c$.

\bigskip

\noindent {\bf Single-trace $T\bar T$-deformed CFTs.} Assuming modular invariance,  we construct the partition function of single-trace $T\bar T$-deformed CFTs, which is found to  be given by 
\eq{
Z_N (\tau, \bar\tau; \mu) & =  \sum_{\{k_1, \dots, k_N\}} \frac{1}{\prod_{n = 1}^N n^{k_n} k_n ! } \prod_{n = 1}^N (T'_n Z)(\tau,  \bar \tau; \mu)^{k_n}, \qquad \sum_{n=1}^N  n k_n = N, \label{intro2}
}
where $Z(\tau,  \bar \tau; \mu)$ is the partition function of the seed $\seedTT$ in $\symTT$, the sum goes over the conjugacy classes of $S_N$, and $T'_n$ is a generalization of the Hecke operator that is defined in \eqref{heckettbar} and matches the one proposed in \cite{Hashimoto:2019hqo}. In addition, we find that  the generating functional of $Z_N (\tau, \bar\tau; \mu)$, see~\eqref{generatingfunctional} and \eqref{generatingfunctional2}, takes the same form as in a symmetric product orbifold CFT,  except that the Hecke operator of the latter is replaced by the generalized Hecke operator $T'_n$.

The partition function \eqref{intro2} allows us to obtain the deformed energies of twisted states in $\symTT$, which are given in terms of the undeformed energies by
\eqsp{
E_L^{(n)}(0) &= E_L^{(n)} (\mu) + \frac{2 \mu}{n} E_L^{(n)}(\mu)  E_R^{(n)}(\mu), \\
 E_R^{(n)}(0) &= E_R^{(n)} (\mu) + \frac{2\mu}{n} E_L^{(n)}(\mu)  E_R^{(n)}(\mu), \label{intro3}
}
where $n$ denotes the twist of the state. These expressions are universal, i.e.~independent of the details of the undeformed symmetric product orbifold $\symCFT$, and match the spectrum of strings winding on TsT-transformed backgrounds \cite{Giveon:2017nie,Apolo:2019zai}. Furthermore, we find that the partition function \eqref{intro2} also matches the partition function of long strings on a linear dilaton background~\cite{Hashimoto:2019hqo}. Hence, our results provide further evidence for the proposed holographic correspondence.

Finally, we show that in the large-$N$ limit, the partition function \eqref{intro2} also satisfies \eqref{intro1}, but in this case it is not necessary to assume the density of light states is sparse. Instead, we show that the density of states at low energies automatically saturates the sparseness bound. This result is analogous to the one found in symmetric product orbifold CFTs \cite{Hartman:2014oaa} and can be understood as a result of the orbifolding together with modular invariance. We expect to find a similarly universal behavior for the torus partition function of other symmetric product orbifolds obtained by single-trace irrelevant deformations, such as those considered in \cite{Chakraborty:2018vja,Apolo:2018qpq,Chakraborty:2019mdf,Apolo:2019yfj,Apolo:2021wcn}.

The paper is organized as follows. In section \ref{se:review} we review the modular invariant properties of $T\bar T$-deformed CFTs, which are used in section \ref{se:partitionfunction} to derive a universal expression for the partition function at large $c$. In that section we propose the sparseness condition for $T\bar T$-deformed CFTs and derive the asymptotic density of states. The partition function of single-trace $T\bar T$-deformed CFTs is derived in section \ref{se:symprod}, where we also derive the spectrum of twisted states. Finally, in section \ref{se:STpartitionfunction} we show that the single-trace $T\bar T$ partition function is universal in the large-$N$ limit and derive its consequences on the density of states.


\section{The partition function of $T\bar T$-deformed CFTs} \label{se:review}

In this section we review two basic properties of $T \bar T$-deformed CFTs: the deformed spectrum and modular invariance of the partition function. In particular, we describe the regime for which the spectrum is well defined and show that modular invariance implies a Hagedorn bound on the temperature of the deformed theory.

Let us consider a $T\bar T$-deformed CFT quantized on a cylinder of size $2\pi R$. For convenience, we set $R = 1$ throughout this paper. The spectrum of the deformed theory can be conveniently written as \cite{Smirnov:2016lqw,Cavaglia:2016oda}
\eqsp{
E_L(0) &= E_L(\mu)  + 2 \mu E_L(\mu) E_R(\mu),  \\
E_R(0) &= E_R(\mu)  + 2 \mu E_L(\mu) E_R(\mu), \label{TTbarspectrum} 
}
where $\mu$ is the deformation parameter and $E_{L,R}(\mu)$ denote the deformed left and right-moving energies. From these expressions, it is not difficult to show that the deformed energy $E(\mu) = E_L(\mu) + E_R(\mu)$ and angular momentum $J(\mu) = E_L(\mu) - E_R(\mu)$ satisfy
\eq{
E(\mu) = - \frac{1}{2\mu} \Big( 1 - \sqrt{1 + {4\mu} E(0) + {4 \mu^2} J(0)^2}\, \Big), \qquad J(\mu) = J(0). \label{TTbarspectrum2}
}
We observe that when $\mu < 0$ the spectrum becomes complex for large values of the undeformed energy $E(0)$. In addition, the energy of the vacuum, which is obtained from \eqref{TTbarspectrum2} by letting $E(0) = -c/12$ and $J(0) = 0$, also becomes complex when $\mu c/3> 1$. For these reasons, in this paper we will assume that the deformation parameter satisfies
\eq{
0 \le \frac{c \mu}{3} \le 1, \label{bound-mu}
}
which guarantees a well-defined spectrum for all values of the undeformed energies. The semiclassical limit of a $T\bar T$-deformed CFT is therefore defined by \cite{Aharony:2018vux}\footnote{For the single-trace version, a large central charge is obtained by taking the number of copies $N$ making up the symmetric product orbifold to be large, without additional restrictions on $\mu$ beyond \eqref{bound-mu}.}
\eq{
c \gg1, \qquad {c \mu} \text{ fixed}. \label{largec}
}

The partition function of a $T\bar T$-deformed CFT on a Euclidean torus is defined in terms of the deformed spectrum by
\eq{
Z(\tau,\bar\tau;\mu) \coloneqq \text{Tr}\Big( q^{ E_L(\mu)} \bar{q}^{E_R( \mu)}\Big) = \sum_{E_L, E_R} \rho(E_L, E_R) e^{2\pi i \tau E_L(\mu) - 2\pi i \bar\tau E_R( \mu)}, \label{Zdef}
}
where $\tau$ is the modular parameter and $q = e^{2\pi i \tau}$. The partition function satisfies the following differential equation~\cite{Cardy:2018sdv}
\eq{
\p_{\mu} Z(\tau, \bar \tau; \mu) = \frac{1}{i \pi} \bigg[ (\tau - \bar \tau) \p_\tau \p_{\bar\tau} - \mu (\p_\tau - \p_{\bar\tau}) \p_{\mu} + \frac{2\mu}{\tau - \bar\tau} \p_{\mu} \bigg] Z(\tau, \bar \tau; \mu), \label{diffeqZ}
}
which is a direct consequence of the differential equation obeyed by the deformed spectrum. Crucially,    the partition function is invariant under modular transformations $\tau \mapsto \frac{a \tau + b}{c\tau + d}$ in the sense that \cite{Datta:2018thy}
\eq{
Z \bigg(\frac{a \tau + b}{c \tau + d} , \frac{a \bar\tau + b}{c \bar\tau + d} ; \frac{\mu}{|c \tau + d|^2} \bigg) = Z (\tau, \bar \tau; \mu).  \label{modularinvariance} 
}
It is important to note that the deformation parameter $\mu$ does not change under modular transformations. Rather, it is the dimensionless deformation parameter $\mu/R^2$ that changes, and the transformation of $\mu$ in \eqref{modularinvariance} comes entirely from the change of the spatial circle, namely $R \mapsto |c \tau + d| R$.

Finally, it is interesting to note that modular $\mathcal S$ transformations ($\tau \mapsto  -1/\tau$), together with the bound \eqref{bound-mu}, imply that
\eq{
|\tau|^2\ge  \frac{c \mu}{3}  \equiv \frac{1}{4\pi^2 T_H^2},  \label{bound-beta}
}
where $T_H$ is the Hagedorn temperature \cite{Giveon:2017nie}. In terms of the deformed temperatures $T_L =1/2\pi \tau$ and $T_R = 1/2\pi \bar\tau$ conjugate to the deformed left and right-moving energies, this bound reproduces the bound on the temperatures of $T\bar T$-deformed CFTs found in \cite{Giveon:2017nie,Apolo:2019zai}. In what follows we will assume that the deformation parameter satisfies \eqref{bound-mu} and that the temperatures lie below the Hagedorn threshold \eqref{bound-beta}.


\section{$T\bar T$ partition functions at large $c$} \label{se:partitionfunction}

In this section we derive universal expressions for the partition functions of $T\bar T$-deformed CFTs when the central charge of the undeformed theory is large. In particular, we will show that, in analogy with the partition function of two-dimensional CFTs at large $c$ \cite{Hartman:2014oaa}, the partition function of $T\bar T$-deformed CFTs is dominated by the vacuum provided that the spectrum of light states is sparse in a way that we will make precise.

\subsection{Review of HKS} \label{se:HKS}

Let us begin by reviewing the universal behavior of the torus partition function of two-dimensional CFTs at large $c$ obtained by HKS \cite{Hartman:2014oaa}. First, it is convenient to perform an analytic continuation of the undeformed partition function and view $Z(\tau,\bar\tau) \coloneqq Z(\tau, \bar\tau, 0)$ as a holomorphic function on $\mathbb C^2$ such that $(\tau, \bar\tau)$ are two independent complex numbers. The modular invariance of the partition function, which holds for $\bar\tau = \tau^*$, can then be shown to hold on $\mathbb C^2$.  

In order to estimate the partition function and extract the density of states, it is convenient to consider the Lorentzian torus that is obtained by setting $(\tau, \bar \tau) = (i\beta_L, -i\beta_R)$ where $\beta_{L,R}$ are two independent and strictly positive real numbers. In terms of $\beta_{L,R}$, invariance of the partition function under modular $\mathcal S$ transformations reads
\be\label{MI-TT-real}
Z(\beta_L,\beta_R) = \text{Tr} \Big( e^{-2\pi\beta_L E_L - 2\pi\beta_R E_R } \Big)  = \text{Tr} \Big( e^{-2\pi\beta'_L E_L -2\pi\beta'_R E_R } \Big) = Z(\beta'_L,\beta'_R), 
\ee
where $\beta'_{L,R} = 1/\beta_{L,R}$. Without loss of generality, modular invariance allows us to restrict the following discussion to the region $\beta_L\beta_R>1$. 

The results of \cite{Hartman:2014oaa} can be summarized as follows. Let us begin with the case where $\beta_{L,R} >1$. Modular invariance of the partition function under $\mathcal S$ transformations \eqref{MI-TT-real} implies that the partition function is dominated by the contribution of the light states, namely
\eq{
\log Z(\beta_L,\beta_R) \approx \log \text{Tr}_L\big(e^{-2\pi\beta_L   E_L-2\pi\beta_R   E_R}\big),
}
where $\approx$ denotes equality up to terms of $o(c)$ and the trace is taken over the light states which satisfy $  E_L +    E_R \leq \epsilon$ for some small positive $\epsilon$. Additionally, if the density of light states is sparse in the sense that
\eq{\label{spa1}
\log\rho(  E_L,  E_R)\lesssim 2\pi\Big(  E_L +   E_R+\frac{c}{12}\Big), \qquad   E_L +   E_R\leq\epsilon,
}
where $\lesssim$ denotes inequality up to corrections of $o(c)$, then the partition function when  $c \gg1$ can be approximated by the contribution of the vacuum
\eq{\label{VD}
\log Z(\beta'_L, \beta'_R) = \log Z(\beta_L, \beta_R) \approx \frac{\pi(\beta_L+\beta_R)c}{12}, \qquad \beta_{L,R}>1.
}
This phenomenon is known as vacuum dominance.

By imposing a stronger sparseness condition on the light states such that
\eq{\label{spa2} 
\log\rho(  E_L,   E_R) \lesssim 4\pi\sqrt{\Big(  E_L + \frac{c}{24}\Big)\Big(  E_R +\frac{c}{24}\Big)}, \qquad   E_L<0 \quad \text{or} \quad   E_R<0,
}
the authors of \cite{Hartman:2014oaa} argued that \eqref{VD}  can be extended to the mixed temperature regime where $\beta_L > 1$, $\beta_R < 1$ or vice versa, with $\beta_L\beta_R>1$. Using modular invariance, the partition function of two-dimensional CFTs at large $c$ can then be shown to satisfy 
\eq{
\log Z(\tau, \bar\tau) \approx \max\Big\{-\frac{\pi i(\tau - \bar\tau )c}{12}, \pi i \Big (\frac{1}{\tau} - \frac{1}{\bar\tau}\Big) \frac{c}{12} \Big\},\qquad |\tau|^2 \neq1, \label{VDfinal}
}
where we switched back to the $(\tau, \bar\tau)$ variables so that $\tau^* = \bar\tau$ and $|\tau|^2 = 1$ is the locus of the Hawking-Page phase transition. An interesting consequence of \eqref{VDfinal} is that it allows us to extend the validity of Cardy's formula for the asymptotic density of states to the semiclassical regime where
\eq{
\log \rho(E_L, E_R) \approx 2\pi \bigg( \sqrt{\frac{c}{6} E_L} + \sqrt{\frac{c}{6} E_R}\, \bigg), \qquad E_L E_R > \bigg(\frac{c}{24}\,\bigg)^2, \quad  c \gg 1 .
}

In the next section, we will show that these arguments can be generalized to $T\bar T$-deformed CFTs modulo small modifications, a fact that will allow us to derive universal expressions for the partition function of $T\bar T$-deformed CFTs at large $c$.


\subsection{Vacuum dominance in $T\bar T$-deformed CFTs} \label{se:partitionTTbar}
 
In this section we consider the torus partition function of $T\bar T$-deformed CFTs. We show that in the large-$c$ limit, modular invariance and a sparse spectrum of light states imply that the ground state dominates the partition function.  As a result, we find that the partition function of $T\bar T$-deformed CFTs is universal at large $c$.

Let us begin by deriving the appropriate sparseness condition for $T\bar T$-deformed CFTs.  In this section we work once again in the regime $\beta_L \beta_R>1$, with the results in the region $\beta_L \beta_R < 1$ obtained by a modular $\mathcal S$ transformation.\footnote{A modular $\mathcal S$ transformation changes the deformation parameter as well, so the discussion for $\beta_L\beta_R<1$ is not perfectly parallel to $\beta_L\beta_R>1$. Nevertheless, as discussed later on, it still suffices to study $\beta_L\beta_R>1$. } We first consider the case where $\beta = \beta_L = \beta_R > 1$. The fact that $E_L(0) = E_R(0) = 0$ is a fixed point of the deformation motivates us to define the light $(L)$ and heavy $(H)$ states of $T\bar T$-deformed CFTs in the same way as in the undeformed CFTs, namely
\eq{
L &\coloneqq \big \{\big(  E_L(\mu),   E_R(\mu) \big) \,|\,   E_L(\mu) +   E_R(\mu) \le \epsilon \big \},\\
H &\coloneqq \big \{\big(  E_L(\mu),   E_R(\mu) \big) \,|\,   E_L(\mu) +   E_R(\mu) > \epsilon \big \}.
}
In addition, it is useful to define
\eq{
Z[L] \coloneqq \text{Tr}_L \, e^{-2\pi\beta E(\mu)} ,\qquad \,\,\, Z[H] &\coloneqq \text{Tr}_H \,e^{-2\pi\beta E(\mu)}, \label{ZLZH}  \\
Z'[L] \coloneqq \text{Tr}_L \, e^{-2\pi\beta' E(\mu')},\qquad Z'[H] &\coloneqq \text{Tr}_H \,e^{-2\pi\beta' E(\mu')}, \label{ZLZHprime}
}
where $  E(\mu) =   E_L(\mu) +   E_R(\mu)$, $\beta' = 1/\beta$, and $\mu' = \mu/\beta^2$. In terms of the light and heavy contributions \eqref{ZLZH} and \eqref{ZLZHprime}, the full partition function of the theory is given by $Z(\beta;\mu) = Z[L] + Z[H] = Z'[L] + Z'[H] = Z(\beta';\mu')$. 

Using \eqref{TTbarspectrum2}, it is not difficult to show that $E(\mu')>E(\mu)$ when $\beta >1$, so that the heavy states $E(\mu) > \epsilon$ satisfy 
\eq{
\beta' E(\mu') - \beta E(\mu) < 0.
}
As a result, we find that $Z[H]$ is bounded by its modular image
\eq{
Z[H]=\text{Tr}_H \big( e^{2\pi\beta' E(\mu')-2\pi\beta E(\mu)}e^{-2\pi\beta' E(\mu')} \big)< \alpha Z'[H], \qquad 0< \alpha < 1. \label{alphadef}
}
This inequality implies that the contribution of the heavy states to the partition function is bounded by the contribution of the light states, namely
\eq{
\frac{1}{\alpha}Z[H] < Z'[H] < Z[H] + Z[L] \quad \Longrightarrow \quad Z[H]<(\alpha^{-1}-1)^{-1}Z[L].
}
Thus, the partition function satisfies
\eq{
\log Z[L] < \log Z(\beta; \mu) < \log Z[L]-\log(1-\alpha), \label{ineqZ}
}
so that in the large-$c$ limit, it is dominated by the contribution of the light states\footnote{Here we have assumed that $\log(1-\alpha)$ is of $\mathcal O(1)$ in the large-$c$ limit.} 
\eq{
\log Z(\beta; \mu) \approx \log \text{Tr}_L\big(e^{-2\pi\beta (E_L(\mu) + E_R(\mu)) }\big). \label{lightstatedom}
}
The inequality \eqref{ineqZ} is similar to the one satisfied by the undeformed theory but with a different value of $\alpha$ \cite{Hartman:2014oaa}.

Once we have established that the light states dominate the partition function at large~$c$, it is not difficult to show that the latter is dominated by the vacuum when the following sparseness condition holds
\eq{ 
\log\rho( E_L , E_R ) \lesssim 2\pi\big[  E_L(\mu)+ E_R(\mu) -  E_{\vac}(\mu) \big], \qquad  E_L(\mu)+ E_R(\mu)\leq\epsilon, \label{spa1-tt}
}
where $  E_{\vac}(\mu)$ is the ground state energy of the $T\bar T$-deformed CFT
\eq{
  E_{\vac}(\mu) = -\frac{1}{2\mu} \bigg(1 - \sqrt{1 - \frac{c \mu}{3}}\,\bigg). \label{vacuumenergy}
}
Note that the energy of the vacuum is large in the large-$c$ limit \eqref{largec}, i.e.~it scales linearly with the central charge, so that it dominates the partition function when \eqref{spa1-tt} holds. Consequently, we find that the partition function of $T\bar T$-deformed CFTs is dominated by the contribution of the vacuum at large $c$, i.e.~
\eq{\label{vd0-tt}
\log Z(\beta;\mu) \approx -2\pi \beta E_{\vac}(\mu), \qquad \beta > 1.
}

The result above is valid when $\beta_L=\beta_R>1$, but it can be easily extended to the $\beta_{L,R}> 1$ region as follows. Since the partition function is greater than the contribution of any single state, the density of states is bounded for any $\beta_{L,R}$ by
\eq{
\rho(E_L, E_R) < Z(\beta_L, \beta_R;\mu) e^{2\pi \beta_L E_L(\mu) + 2\pi \beta_R E_R(\mu)}.\label{rhobound}
}
Then, \eqref{vd0-tt} implies that $\rho(E_L,E_R)<e^{2\pi [E_L(\mu)+ E_R(\mu) -  E_{\vac}(\mu)]}$, such that when $\beta_{L,R} > 1$, the partition function satisfies
\eq{
Z(\beta_L,\beta_R;\mu) = \sum_{E_L, E_R} \rho(E_L,E_R)e^{-2\pi\beta_LE_L(\mu)-2\pi\beta_RE_R(\mu)} < \alpha' e^{-\pi(\beta_L + \beta_R) E_{\vac}(\mu)},
}
where $\alpha'$ is a numerical constant that is independent of $c$. Since the partition function cannot be smaller than the contribution of the vacuum, and $\alpha'$ is independent of $c$, we see that the partition function is still dominated by the vacuum when $\beta_{L, R} > 1$, namely
\eq{
\log Z(\beta_L, \beta_R;\mu) \approx -\pi (\beta_L + \beta_R) E_{\vac}(\mu), \qquad \beta_{L,R} > 1.
}

In order to extend vacuum dominance to the mixed temperature regime where $\beta_L > 1$, $\beta_R < 1$ or vice versa, it is necessary to impose a stronger sparseness condition on the light states. In this case, the light states are defined in analogy with \cite{Hartman:2014oaa} by
\eq{
E_L(\mu) < 0 \quad \textrm{or} \quad E_R (\mu) < 0.
}
In contrast, the heavy states satisfy $E_{L,R}(\mu) > 0$. The appropriate sparseness condition in the mixed temperature regime can be reverse engineered from \eqref{rhobound} by assuming that the partition function is dominated by the vacuum. It follows that the density of states is bounded by
\eq{
\rho(E_L,E_R) < e^{ 2\pi \beta_L E_L(\mu) + 2\pi \beta_R E_R(\mu) -  \pi (\beta_L + \beta_R) E_{\vac}(\mu)}.
}
Optimizing the exponent over all $\beta_{L,R}$ satisfying $\beta_L\beta_R>1$, we find that it suffices to impose the sparseness condition
\eq{
\log\rho ( E_L , E_R ) \lesssim 4\pi\sqrt{\bigg( E_L(\mu) - \frac{1}{2}  E_{\vac}(\mu)\bigg)\bigg( E_R(\mu) -  \frac{1}{2} E_{\vac}(\mu)\bigg)}, \label{spa2-tt}
}
which reduces to the sparseness condition of the undeformed CFT \eqref{spa2} in the limit $\mu \to 0$.

It is interesting to note that the sparseness conditions on the density of light states \eqref{spa1-tt} and \eqref{spa2-tt}, which are written in terms of the deformed spectrum, take the same form as those of the undeformed CFT, namely \eqref{spa1} and \eqref{spa2}. However, the former differ from the latter when \eqref{spa1-tt} and \eqref{spa2-tt} are rewritten in terms of the undeformed energies using~\eqref{TTbarspectrum}. In other words, the sparseness conditions in $T\bar T$-deformed CFTs cannot be obtained by rewriting those of the undeformed CFT in terms of the  deformed quantities, and hence the bound takes different numerical values. This situation contrasts with the density of high energy states described later, which takes the same numerical value before and after the deformation, but differs in its functional form: it scales as $\log\rho  \propto \sqrt{E(0)}$ at large $E(0)$ before the deformation but as $\log\rho\propto E(\mu)$ after the deformation. 

It is also worth noting that the upper bounds on the density of light states in \eqref{spa1-tt} and \eqref{spa2-tt} are increasing functions of $\mu$. As a result, the sparseness conditions of $T\bar T$-deformed CFTs become weaker as $\mu$ grows. This means, in particular, that a sparse CFT remains sparse after turning on the $T\bar T$ deformation. On the other hand, a sparse $T\bar T$-deformed CFT does not imply that the undeformed CFT is sparse.  Additionally, recall that in order to approximate the partition function in the high temperature regime $\beta_L \beta_R < 1$, we need to perform a modular $\mathcal S$ transformation and consider vacuum dominance at deformation parameter $\mu'  = \mu/ \beta_L \beta_R > \mu$. Since the sparseness conditions grow weaker as $\mu$ increases, vacuum dominance at $\mu'$ is implied by vacuum dominance at $\mu$. Therefore, for a given $\mu$, it suffices to consider the case $\beta_L\beta_R>1$. 

Using the stronger sparseness condition \eqref{spa2-tt}, the argument of vacuum dominance in the region $\beta_L \beta_R > 1$ proceeds as in CFT, provided that we keep track of the change in $\mu$ under modular transformations. As a result, we find that the partition function of sparse $T\bar T$-deformed CFTs is universal in the large-$c$ limit and is given by
\eq{
\log Z(\tau, \bar\tau; \mu) \approx \max \big\{\pi i (\tau - \bar\tau) E_{\vac}(\mu), -\pi i \Big (\frac{1}{\tau} - \frac{1}{\bar\tau}\Big) E_{\vac}\Big(\frac{\mu}{\tau\bar\tau}\Big) \big\}, \qquad |\tau|^2 \ne 1, \label{approxZttbar}
}
where $  E_{\vac}(\mu)$ is the ground state energy defined in \eqref{vacuumenergy}. In analogy with the partition function of the undeformed CFT, the $T\bar T$-deformed partition function is dominated by the vacuum when $|\tau|^2 > 1$ and by its $\mathcal S$ modular image when $|\tau|^2 < 1$. In the latter case, we must keep track of the transformation of $\mu$ in the definition of the ground state energy, which has important repercussions for the asymptotic density of states, as we now describe.


\subsection{Asymptotic density of states in the semiclassical limit} \label{se:entropy}
 
Let us now use the large-$c$ partition function of $T\bar T$-deformed CFTs obtained in the previous section to derive the asymptotic density of states in the canonical and microcanonical ensembles. 
 
In the canonical ensemble, the logarithm of the density of states can be obtained from the approximate partition function \eqref{approxZttbar} in the region $|\tau|^2 < 1$ via
\eq{
S  &=(1- \tau\p_\tau-{\bar\tau}\p_{\bar\tau})\log Z(\tau,\bar{\tau} ; \mu) \approx \frac{i\pi c}{6} \frac{1}{\g} \bigg( \frac{1}{\tau} - \frac{1}{\bar{\tau}} \bigg), \qquad \gamma = \sqrt{1 - \frac{\mu c}{3 |\tau|^2}}. \label{ttbarentropy}
}
We can express the entropy in the microcanonical ensemble by trading the inverse temperatures for the thermal expectation values of the energies in the usual way, namely
\eq{
  E_L(\mu)  &=  \frac{1}{2\pi i}  \p_\tau \ln Z(\tau,\bar{\tau} ; \mu)= \frac{1}{2}\bigg(\frac{1}{\tau^2} + \frac{\tau - \bar\tau}{|\tau|^2} \p_\tau\bigg)   E_{\vac}\Big(\frac{\mu}{\tau\bar\tau}\Big) ,  \\
  E_R(\mu)   &= - \frac{1}{2\pi i} \p_{\bar\tau} \ln Z(\tau,\bar{\tau} ;\mu)= \frac{1}{2}\bigg(\frac{1}{\bar\tau^2} - \frac{\tau - \bar\tau}{|\tau|^2} \p_{\bar\tau}\bigg)   E_{\vac} \Big(\frac{\mu}{\tau\bar\tau}\Big).
}
After some algebra, these equations can be shown to satisfy
  \eqsp{
  \tau = & \frac{i}{2} \Bigg( \sqrt{\frac{\frac{c}{6}(1 + 2 \mu   E_R(\mu))}{  E_L(\mu)}} + 2\mu \sqrt{\frac{\frac{c}{6}   E_R(\mu)}{1+ 2\mu  E_L(\mu)}}\, \Bigg), \\
    \bar\tau = - & \frac{i}{2} \Bigg( \sqrt{\frac{\frac{c}{6}(1 + 2 \mu   E_L(\mu))}{  E_R(\mu)}} + 2\mu \sqrt{\frac{\frac{c}{6}   E_L(\mu)}{1+ 2\mu  E_R(\mu)}}\, \Bigg). \label{tautaubarsol}
  }
 
Using \eqref{tautaubarsol} we can write the entropy \eqref{ttbarentropy} in terms of the thermal expectation values of the energies such that in the microcanonical ensemble it reads\footnote{Alternatively, we can obtain \eqref{ttbarentropy2} by taking the inverse Laplace transform of the partition function.}
\eq{
S  & \approx 2\pi \bigg \{ \sqrt{\frac{c}{6} E_L(\mu) \Big(1 +  {2\mu} E_R(\mu) \Big)} + \sqrt{\frac{c}{6} E_R(\mu) \Big( 1 + {2\mu} E_L(\mu) \Big) }\, \bigg \}. \label{ttbarentropy2}
}
Note that \eqref{ttbarentropy} is valid at high temperatures where $|\tau|^2 < 1$,  which means that in the microcanonical ensemble the entropy is given by \eqref{ttbarentropy2} for energies satisfying \eq{ \label{cardyregime}
\frac{ E_L( \mu) E_R( \mu)}{1 +  {2\mu} \big ( E_L( \mu)  +  E_R( \mu)\big )} > \frac{ E_{\vac}(\mu) ^2 }{4\big(1 + {2\mu} E_{\vac}(\mu)\big) }.
}
Eq.~\eqref{cardyregime} tells us that the left and right-moving energies must be greater than the central charge up to corrections that depend on $c\mu$ and therefore remain fixed in the semiclassical limit, namely
\eq{
\frac{E_L(\mu)}{c} \frac{E_R(\mu)}{c} > \frac{1}{24^2}\big( 1 + \mathcal O(c\mu)\big).
}
Thus, in analogy with HKS, we find that there is a set of states with energies lying between the light states and \eqref{cardyregime}, for which the partition function is not universal and the density of states is sensitive to the details of the theory. In the nomenclature of \cite{Hartman:2014oaa}, these states correspond to the $T\bar T$-deformed analogs of the enigmatic states of two-dimensional CFTs.


\section{Symmetric product orbifolds of $T\bar T$-deformed CFTs} \label{se:symprod}

In this section we consider the single-trace deformation of a symmetric product orbifold CFT, which is equivalent to the symmetric product orbifold of the $T\bar T$-deformed CFT. We will show that the orbifolding procedure, together with modular invariance, imply the existence of additional twisted states in the spectrum of the deformed theory. In addition, we construct the modular invariant partition function and show that the energy of the twisted states matches the spectrum of winding strings on the TsT-transformed backgrounds of \cite{Giveon:2017nie, Apolo:2019zai}.

\subsection{Symmetric product orbifolds and modular invariance}

Let us consider a $T\bar T$-deformed CFT that we denote by $\seedTT$ and whose action satisfies the differential equation \eqref{TTbardefinition}. The symmetric product orbifold of this theory (or symmetric orbifold for short)  is defined by
\eq{
\symTT \coloneqq \frac{(\seedTT)^N}{S_N}, \label{symTTbar}
}
where $N$ is a positive integer and $S_N$ is the symmetric group. The symmetric orbifold~\eqref{symTTbar} can be interpreted as a ``single-trace'' deformation of the symmetric orbifold of the undeformed CFT ($\symCFT$) in the sense that the action $I_N$ of \eqref{symTTbar} satisfies
\eq{
\frac{\p I_N}{\p \mu}= 8\pi  \sum_{i=1}^N \int d^2 x \,  \, (T\bar T)^{(i)}, \qquad (T\bar T)^{(i)} \coloneqq \frac{1}{8} \big( T^{(i)\a\b} T^{(i)}_{\a\b} - (T_{\a}^{(i)\a})^2 \big) , \label{symTTbaraction}
}
where $T_{\mu\nu}^{(i)}$ denotes the components of the stress tensor in the $i$th copy of the seed $\seedTT$. In contrast, the double-trace deformation of $\symCFT$ contains additional terms on the right-hand side of \eqref{symTTbaraction} that mix the stress tensors on different copies of the seed theory so that the action satisfies \eqref{TTbardefinition} instead of \eqref{symTTbaraction}. In principle, the single-trace deformation can be applied to other product theories, such as permutation orbifolds, and it would be interesting to study such families of $T\bar T$-deformed theories.  

We are interested in the partition function of $\symTT$ which is defined by
\eq{
Z_N (\tau, \bar \tau; \mu) \coloneqq \textrm{Tr}\Big( q^{E_L(\mu)} \bar q^{E_R(\mu)} \Big). \label{ZNdef}
}
Since the partition function $Z(\tau, \bar\tau,; \mu)$ of the deformed seed $\seedTT$ is modular invariant \eqref{modularinvariance}, it is natural to expect the partition function of $\symTT $ to be modular invariant as well, namely we expect it to satisfy
\eq{
Z_N \bigg(\frac{a \tau + b}{c \tau + d} , \frac{a \bar\tau + b}{c \bar\tau + d} ; \frac{\mu}{|c \tau + d|^2} \bigg) = Z_N (\tau, \bar \tau; \mu). \label{modularinvariance2}
}
This is in analogous to the discussion of symmetric product CFTs, where twisted states can be added to guarantee modular invariance of the partition function after the symmetrization procedure \cite{Klemm:1990df}. This expectation raises a puzzle, however, since \cite{Aharony:2018bad} has shown that a modular invariant partition function satisfying \eqref{modularinvariance2} implies that the spectrum of the theory satisfies \eqref{TTbarspectrum2}. This spectrum corresponds to the double-trace $T\bar T$ deformation of $\symCFT$, instead of the single-trace deformation considered in this section.\footnote{In order to see that the spectrum of the double and single-trace theories is different, note that the energy of a state in the single-trace version receives contributions from the states in each copy of $\symTT$. Consequently, while the double-trace spectrum consists of square roots \eqref{TTbarspectrum2}, the single-trace spectrum contains sums of square roots which cannot be obtained from \eqref{TTbarspectrum2}.}

The resolution of this puzzle lies on the assumptions made in \cite{Aharony:2018bad}, which assumes that there is only one parameter in the theory such that the energy $E_m(0)$ of any state $\Ket{m}$ in the undeformed CFT is deformed uniquely to the energy $E_m(\mu)$ of a deformed state $\Ket{m}_\mu$ such that
\eq{
E_m(0) \mapsto E_m(\mu)= f (E_m(0), J_m(0), \mu), \qquad J_m(0) \mapsto J_m(\mu).
}
This assumption is not valid for the single-trace deformation of a symmetric orbifold where the map between the undeformed and deformed energies depends on the distribution of the energy into the different copies of $\symTT$. In this case, the states are labelled by additional parameters $\{m_i\}$ so that the total energy is a sum $E_m = \sum_{i} E_{m_i}$, and we have 
\eq{
E_{m}(0) \mapsto E_m(\mu)=\tilde f (E_{m_i}(0), J_{m_i}(0), \mu; \{m_i\}), \qquad J_m(0) \mapsto J_m(\mu). \label{defmap}
}
Crucially, the undeformed symmetric orbifold features degenerate states that are lifted by the deformation such that the map \eqref{defmap} is no longer one-to-one. It follows that the results of \cite{Aharony:2018bad} do not apply to single-trace $T\bar T$ deformations so that modular invariance of the partition function does not imply that the spectrum is given by~\eqref{TTbarspectrum2}, in line with the structure of the spectrum of $\symTT$. Another way of saying this is that the partition function of $\symTT$ is assumed to be modular invariant but it does not to satisfy the differential equation \eqref{diffeqZ}.

To summarize, the symmetric orbifold of a $T\bar T$-deformed CFT is defined by \eqref{symTTbar} and its partition function is assumed to be modular invariant \eqref{modularinvariance2}. In what follows we will use modular invariance to construct the partition function of $\symTT$ and derive the spectrum of twisted states.


\subsection{The single-trace $T\bar T$ partition function}

Let us now derive the partition function of single-trace $T\bar T$-deformed CFTs. In analogy with symmetric orbifolds of two-dimensional CFTs, the spectrum of $\symTT$ consists of untwisted states obtained from the symmetrized product of $N$ copies of the seed theory, as well as twisted states, which are required to preserve the modular invariance of the partition function. In order to see the emergence of the twisted states and determine their energies, we must first consider the contribution of the untwisted states to the partition function.

The untwisted sector of $\symTT$ can be obtained from the product of states $\phi^{(i_n)}$ from each copy of $\seedTT$ in the product theory $(\seedTT)^N$, i.e.~from $\otimes_{n=1}^N \phi^{(i_n)}$, where $n$ labels different copies of the symmetric product and $i_n$ labels the state on the $n$th copy. Any such configuration where at least two of the $\phi^{(i_n)}$ states are distinct is not invariant under the action of the symmetric group. As a result, the orbifolding procedure leads to a reduction in the number of states and the physical states of the theory take the form 
\eq{
\Phi = \textrm{Sym}(\otimes_{n=1}^N \phi^{(i_n)}). \label{productstates}
}
The contribution of generic states (where all of the $\phi^{(i_n)}$ are distinct) to the partition function of $\symTT$ can be obtained from $\frac{1}{N!}Z(\tau, \bar \tau;\mu)^N$, where $Z(\tau, \bar\tau;\mu)$ is the partition function of the seed theory. This normalization miscounts several states, however, since there are states in $Z(\tau, \bar \tau;\mu)^N$ that are invariant under elements of $S_N$. The simplest example of such states is $\otimes_{n=1}^N \phi^{(1)}$ --- where $\phi^{(1)}$ is some state in $\seedTT$ --- which is invariant under any element of the symmetric group. In order to account for these kind of states we can either subtract them from $Z(\tau, \bar \tau;\mu)^N$ before normalization, or add additional contributions to $\frac{1}{N!}Z(\tau, \bar \tau;\mu)^N$ which reproduce the appropriate normalization of these states.

Let us illustrate how we can obtain the partition function for the untwisted sector of $\symTT$ in the simple case where $N = 3$.\footnote{The case $N = 2$ is too simple as it corresponds to the $\mathbb Z_2$ orbifold of the seed $\seedTT$.} This example will provide some intuition for the general formula reported below, which relies only on the structure of $S_N$, and is hence universally valid for any symmetric orbifold. 

The untwisted sector of the symmetric orbifold $\symTT$ admits the following kinds of states when $N = 3$,
\eqsp{
\Phi_{(i)} &\coloneqq   \phi^{(i)} \otimes \phi^{(i)} \otimes \phi^{(i)},  \\
\Phi_{(i,j)} &\coloneqq \textrm{Sym}(\phi^{(i)} \otimes \phi^{(i)} \otimes \phi^{(j)}), \qquad  i\ne j, \\
\Phi_{(i,j,k)} &\coloneqq \textrm{Sym}(\phi^{(i)} \otimes \phi^{(j)} \otimes \phi^{(k)}), \qquad i \ne j \ne k,\label{statesp3}
}
where $\Phi_{(i)}$ consists of three copies of the same field $\phi^{(i)}$ --- one from each copy of $(\seedTT)^3$, $\Phi_{(i,j)}$ contains two copies of $\phi^{(i)}$ and one copy of $\phi^{(j)}$, and $\Phi_{(i,j,k)}$ consists of different fields from each copy of the product theory. Since all of the copies in $\Phi_{(i)}$ are the same, its contribution to the partition function is equivalent to multiplying the modular parameter by a factor of $3$, namely $Z(3\tau, 3\bar\tau; \mu)$. Similarly, we find that the contribution of each of these kinds of states to the partition function of $\textrm{Sym}^3 \seedTT$ is
\eqsp{
\Phi_{(i)} &:  \quad Z^{(3)}(\tau, \bar\tau; \mu) \coloneqq Z(3\tau, 3\bar\tau; \mu) ,    \\
\Phi_{(i,j)} &: \quad Z^{(2)}(\tau, \bar\tau; \mu) \coloneqq Z(2\tau, 2\bar\tau; \mu) Z(\tau, \bar\tau; \mu) - Z^{(3)}(\tau, \bar\tau; \mu) ,  \\
\Phi_{(i,j,k)} &: \quad Z^{(1)}(\tau, \bar\tau; \mu) \coloneqq \tfrac{1}{3!} \big[ Z(\tau, \bar\tau; \mu)^3 - 3  Z^{(2)}(\tau, \bar\tau; \mu) -  Z^{(3)}(\tau, \bar\tau; \mu) \big],\label{partitionsp3}
}
where in the second and third lines we subtracted the contributions of the $\Phi_{(i)}$ and $\Phi_{(i,j)}$ states from $Z(2\tau, 2\bar\tau; \mu) Z(\tau, \bar\tau; \mu)$ and $Z(\tau, \bar\tau; \mu)^3$. Adding together the terms in \eqref{partitionsp3}  we find that the contribution of the untwisted states to the partition function of $\textrm{Sym}^3 \seedTT$ is simply given by
\eq{
Z_{\textrm{untwisted}}(\tau, \bar\tau; \mu) & = \tfrac{1}{3!} \big[ Z(\tau, \bar\tau; \mu)^3  + 3 Z(2\tau, 2\bar\tau; \mu) Z(\tau, \bar\tau; \mu) + 2  Z(3\tau, 3\bar\tau; \mu) \big]. \label{untwistedZp3}
}

We can interpret the untwisted sector partition function \eqref{untwistedZp3} in the following way. Let us associate the partition functions $Z(\tau,\bar\tau;\mu)$, $Z(2\tau, 2\bar\tau;\mu)$, and $Z(3\tau,3 \bar\tau;\mu)$ with the $\mathbb I$, $\mathbb Z_2$, and $\mathbb Z_3$ cycles of the symmetric group, respectively, where $\mathbb I$ denotes the trivial cycle, i.e.~the identity. The partition function \eqref{untwistedZp3} is then seen to correspond to the cycle index of $S_3$ where, modulo the overall normalization, each of the coefficients in \eqref{untwistedZp3} is the size of the corresponding conjugacy class of $S_3$.

For arbitrary values of $N$, the contribution of the untwisted sector of $\symTT$ to the partition function can be obtained in a similar way from the cycle index of $S_N$. Let us identify $Z(n\tau, n\bar\tau;\mu)$, $n > 1$ with the $\mathbb Z_n$ cycle of $S_N$. The partition function for the untwisted sector of $\symTT$ can then be written as \cite{Dijkgraaf:1996xw,Dijkgraaf:1998zd,Bantay:2000eq}
\eq{
Z_{\textrm{untwisted}}(\tau, \bar\tau; \mu) & =  \sum_{\{k_1, \dots, k_N\}} \frac{1}{\prod_{n = 1}^N n^{k_n} k_n ! } \prod_{n = 1}^N Z(n \tau,  n \bar \tau; \mu)^{k_n}, \label{untwistedZ}
}
where $\{k_1, k_2, \dots, k_N\}$ labels the conjugacy classes of $S_N$ with $k_n$ the number of $\mathbb Z_n$ cycles in each conjugacy class. The $k_n$ numbers are constrained to satisfy 
\eq{
\sum_{n=1}^N  n k_n = N.
}
Note that each factor of $n$ in the denominator of \eqref{untwistedZ} counts the size of each $\mathbb Z_n$ while $k_n!$ counts the permutations of the $\mathbb Z_n$ cycles. 
For the $N = 3$ case considered above, the conjugacy classes of $S_3$ are denoted by $\{3, 0,0\}$, $\{1,1,0\}$, $\{0,0,1\}$, and it is not difficult to verify that \eqref{untwistedZ} reproduces \eqref{untwistedZp3}.

The partition function \eqref{untwistedZ} is not modular invariant because of each of the $Z(n\tau, n\bar\tau; \mu)$ terms associated with the non-trivial ($n \ne 1$) conjugacy classes of $S_N$. Indeed, although $Z (n\tau, n\bar \tau;\mu)$ is invariant under $\mathcal T$ transformations ($\tau \mapsto \tau + 1)$, it fails to be invariant under $\mathcal S$ transformations ($\tau \mapsto  -1/\tau$) since
\eq{
\mathcal S \cdot Z (n\tau, n\bar \tau;\mu) & = Z  \bigg(\!\!-\frac{n}{\tau}, -\frac{n}{\bar\tau}; \frac{\mu}{|\tau|^2}\bigg) = Z \bigg(\frac{\tau}{n}, \frac{\bar \tau}{n};\frac{\mu}{n^2} \bigg), \label{S1}
}
where we used the modular invariance of $Z (\tau, \bar \tau;\mu)$. In order to construct a modular invariant partition function we need to add additional terms to the partition function \eqref{untwistedZ} which give rise to the twisted states characteristic of orbifold theories.

Let us begin by assuming that $n$ is prime. We follow closely the approach of \cite{Klemm:1990df}, where modular invariance was used to derive the spectrum of twisted states in $\mathbb Z_n$ orbifolds of two-dimensional CFTs. We can make each of the $Z (n\tau, n\bar \tau;\mu)$ functions in \eqref{untwistedZ} invariant under $\mathcal S$ transformations by adding the modular image \eqref{S1}. This term spoils modular invariance under $\mathcal T$ transformations, however, since for any integers $k$ and $\a \in [1,\, n-1]$ we have
\eq{
 \mathcal T^{\a + k n} \cdot Z \bigg(\frac{\tau}{n}, \frac{\bar \tau}{n};\frac{\mu}{n^2} \bigg) = Z \bigg(\frac{\tau + \a}{n}, \frac{\bar \tau +\a}{n};\frac{\mu}{n^2} \bigg). \label{TS1}
}
Performing an additional $\mathcal S$ transformation of \eqref{TS1} we observe that
\eq{
\mathcal S \cdot \mathcal T^{\a + k n} \cdot & Z \bigg(\frac{\tau}{n}, \frac{\bar \tau}{n};\frac{\mu}{n^2} \bigg)  = Z \bigg(\frac{\frac{\a\tau}{n} - \frac{1}{n}}{\tau}, \frac{\frac{\a\bar\tau}{n} - \frac{1}{n}}{\bar\tau};\frac{\frac{\mu}{n^2}}{|\tau|^2} \bigg)  \notag \\
&\hspace{55pt} =  (\widetilde{\mathcal T})^{\frac{\tilde\alpha}{ n}} \cdot Z \bigg(\frac{ \a\tilde\tau - \frac{1}{n} (1+ \a \tilde \a)}{n \tilde\tau - \tilde \a}, \frac{ \a\bar{\tilde\tau}  -   \frac{1}{n} (1+ \a \tilde \a)}{n\bar{\tilde\tau} - \tilde \a} ;\frac{\tilde\mu}{|n\tilde\tau - \tilde \alpha|^2} \bigg),\label{STS1}
}
where $\tilde\tau = \frac{\tau}{n}$, $\bar{\tilde\tau} = \frac{\bar\tau}{n}$, $\tilde\mu = \frac{\mu}{n^2}$, $\tilde \alpha$ is a positive integer, and $\widetilde{\mathcal T}$ is a $\mathcal T$ transformation with respect to $\tilde\tau$. 

The argument of the partition function in the second line of \eqref{STS1} can be written as an $SL(2,\mathbb R)$ transformation with respect to $\tilde\tau$, namely $Z\big(\tfrac{a\tilde\tau + b }{c \tilde\tau + d},\tfrac{a\bar{\tilde\tau} + b }{c \bar{\tilde\tau} + d}; \frac{\tilde\mu}{|c\tilde\tau + d|^2}\big )$, where the $a$, $b$, $c$, and $d$ parameters satisfy $ad - bc = 1$ and are given by
\eq{
a = \a, \qquad b = -\frac{1}{n} (1+ \a \tilde \a), \qquad c = n, \qquad d = -  \tilde \a,\qquad \alpha,\, \tilde\alpha, \,n \in \mathbb Z.
}
This transformation will correspond to a modular $SL(2,\mathbb Z)$ transformation provided that $b$ is an integer. When $n$ is prime,  $\alpha\in[1,\,n-1]$ is co-prime with $n$ and Bezout's identity tells us that it is always possible to find a pair of integers $(\til\a,\,\til k)$ such that 
\eq{
\a \til\a+1= \tilde{k} n.\label{bezout}
}
It follows that $b$ is an integer. Furthermore, we note that given any $\a \in [1, \, n - 1]$, there is a unique integer $\til\a \in [1, \, n - 1]$ satisfying \eqref{bezout}, and conversely given any $\til\a \in [1, \, n - 1]$ there is also a unique integer $\a \in [1, \, n - 1]$. In order words, there is a one-to-one map between the integers $\alpha$ and $\tilde\alpha$ if both are restricted to the region $[1, \, n - 1]$.  Thus, we can use the modular invariance of $Z(\tau, \bar\tau; \mu)$ to write \eqref{STS1} in the following way
\eq{
\mathcal S \cdot \mathcal T^{\a + k n} \cdot Z \bigg(\frac{\tau}{n}, \frac{\bar \tau}{n};\frac{\mu}{n^2} \bigg) & =  \big(\widetilde{\mathcal T}\big)^{\frac{\til\a}{n}} \cdot Z (\tilde\tau, \bar{\tilde\tau}; \tilde\mu) = \mathcal T^{\til\a + k n}\cdot Z \bigg(\frac{\tau }{n}, \frac{\bar \tau }{n};\frac{\mu}{n^2} \bigg), \label{STS2}
}
where in the last equality we have added $\mathcal T^{k n}$ since it leaves $Z \big(\frac{\tau }{n}, \frac{\bar \tau }{n};\frac{\mu}{n^2} \big)$ invariant.

We have shown that the $\mathcal S$ transformation of \eqref{TS1} yields another partition function \eqref{STS2} which takes the same form as the left hand side of \eqref{TS1}, but with the integer $\alpha\in[1,\,n-1]$ replaced by another integer $\tilde\alpha\in[1,\,n-1]$. As a result,  adding \eqref{S1} and \eqref{TS1} to $Z (n\tau, n\bar \tau;\mu)$ for all of the possible values of $\a \in [1,\,  n-1]$ is sufficient to render $Z (n\tau, n\bar \tau;\mu)$ invariant under any combination of $\mathcal T$ and $\mathcal S$ transformations, and hence invariant under any modular transformation.  Therefore, when $n$ is prime, the following linear combination of modular images of $Z (n \tau, n \bar \tau;\mu)$ is modular invariant
\eq{
Z (n \tau, n \bar \tau;\mu) + \sum_{\a = 0}^{n -1}  Z \bigg( \frac{\tau + \a}{n}, \frac{\bar \tau + \a}{n};\frac{\mu}{n^2} \bigg) . \label{modularZn}
} 

In order to discuss the case for general $n$, it is useful to relate \eqref{modularZn} to the action of a generalized version of the Hecke operator on the seed partition function $Z( \tau,  \bar \tau;\mu)$. When $n$ is prime, the action of the ordinary Hecke operator $T'_n$ on a modular form $f(\tau)$ of weight 0 is defined by (see e.g.~\cite{Apostol:1990})\footnote{Our convention of the Hecke operator differs from the standard one $T_n$ by an overall rescaling, $T'_n = n T_n$.}
\eq{
(T'_n f)(\tau) = f( n\tau) + \sum_{\a = 0}^{n-1} f\bigg(\frac{\tau + \a}{n}\bigg), \qquad n \in \mathbb P; \label{hecke0}
}
and more generally, when $n$ is an arbitrary positive integer, by
\eq{
(T'_n f)(\tau)  = \sum_{\gamma | n} \sum_{\a = 0}^{\gamma-1} f\bigg(\frac{n \tau + \a \gamma}{\gamma^2}\bigg), \qquad n \in \mathbb Z^+. \label{hecke1}
}
The similarity between  \eqref{modularZn} and \eqref{hecke0} suggests that we can interpret \eqref{modularZn} as the action of the Hecke operator on $Z(\tau, \bar \tau; \mu)$, which has been generalized to include a rescaling of the deformation parameter $\mu$ by a factor of $n^{-2}$. It is then natural to generalize the Hecke operator \eqref{hecke1} so that its action on the partition function of $T\bar T$-deformed CFTs is given, for any positive integer $n$, by
\eq{
(T'_n Z) ( \tau,  \bar \tau; \mu) =  \sum_{\gamma | n} \sum_{\a = 0}^{\gamma-1} Z \bigg( \frac{n \tau + \a \gamma}{\gamma^2}, \frac{n \bar \tau + \a \gamma }{\gamma^2};\frac{\mu}{\gamma^2} \bigg). \label{heckettbar}
}
The generalized Hecke operator \eqref{heckettbar} reproduces the modular invariant expression \eqref{modularZn} when $n$ is prime, and furthermore matches the operator obtained holographically from the worldsheet theory of strings on a linear dilaton background \cite{Hashimoto:2019hqo}.

Let us now prove that \eqref{heckettbar} is indeed the appropriate generalization of the Hecke operator by showing that $(T'_n Z)(\tau,\bar\tau;\mu)$ is modular invariant for any positive integer $n$. We will follow closely the proof given in \cite{Apostol:1990} for the modular invariance of \eqref{hecke1}, and show that the scaling of $\mu$ in \eqref{heckettbar} guarantees that $(T'_n Z) ( \tau,  \bar \tau; \mu) $ remains modular invariant. The essence of the proof lies on the fact that the Hecke transform of $Z(\tau, \bar\tau; \mu)$ can be equivalently written as 
\eq{
(T'_n Z) ( \tau,  \bar \tau; \mu) =  \sum_{A_1}\, Z \Big(A_1\tau, A_1\bar\tau; \frac{\rho_1^2}{n^2}  \mu \Big), \label{heckettbar2}
}
where we sum over a complete set of inequivalent elements of $\Gamma(n)$ which can be parametrized by the upper triangular matrices\footnote{Here $\Gamma(n)$ denotes the set of all transformations $\tau \mapsto M \tau = \frac{\rho \tau + \s}{\eta \tau + \d}$ with $\rho\d - \eta\s = n$ and $\rho,\s,\eta,\d \in \mathbb Z$.}
\renewcommand{\arraystretch}{1.2}
\eq{
A_1 = \left( \begin{array}{cc}
\rho_1 & \s_1 \\
0 & \d_1 
\end{array} \right), \qquad \rho_1 \d_1 = n, \quad  \s_1 \in \mathbb Z \,(\text{mod } \d_1). \label{Adef}
}
It is not difficult to verify that the properties of $\rho_1$, $\s_1$, and $\d_1$ given above imply that \eqref{heckettbar2} is equivalent to \eqref{heckettbar} with $(\rho_1,\s_1,\d_1) = (n/\gamma,\alpha,\gamma)$. 

Theorem 6.9 of \cite{Apostol:1990} states that, given $A_1 \in \Gamma(n)$ and a standard modular transformation represented by the matrix $M_1 \in \Gamma(1)$, it is always possible to find $M_2 \in \Gamma(1)$ and an upper triangular  matrix $A_2  \in \Gamma(n)$ such that
\eq{
A_1 M_1 = M_2 A_2, \label{theorem1}
}
where the map is one-to-one. Furthermore, if we parametrize the $A_i$ and $M_i$ matrices by
\eq{
A_i = \left( \begin{array}{cc}
\rho_i & \s_i \\
0 & \d_i 
\end{array} \right), \quad 
M_i = \left( \begin{array}{cc}
a_i & b_i \\
c_i & d_i 
\end{array} \right), \qquad i = 1, 2,
}
where $\rho_2$, $\s_2$, and $\d_2$ satisfy the same properties as $\rho_1$, $\s_1$, and $\d_1$ in \eqref{Adef}, then it is not difficult to show that
\eq{
\frac{\rho_1^2}{ |c_1 \tau + d_1|^2} = \frac{\rho_2^2}{|c_2 A_2\tau + d_2|^2}. \label{theorem2}
}

The properties of $A_i \in \Gamma(n)$ described above imply that, under a modular transformation $\tau \mapsto M_1 \tau = \frac{a_1 \tau + b_1}{c_1 \tau + d_1}$, the Hecke-transformed partition function \eqref{heckettbar2} satisfies
\eq{
\!\!\!\!\!\!\!\!\!\!\!\!\!\! (T'_n Z)\bigg(\frac{a_1 \tau + b_1}{c_1 \tau + d_1} , \frac{a_1 \bar\tau + b_1}{c_1 \bar\tau + d_1} ; \frac{\mu}{|c_1 \tau + d_1|^2} \bigg)  &=  \sum_{A_1}\, Z \Big(A_1 M_1 \tau, A_1M_1\bar\tau;   \frac{\rho_1^2}{|c_1 \tau + d_1|^2} \frac{\mu}{ n^2}  \Big) \notag\\
&=  \sum_{A_2}\, Z \Big(M_2 A_2 \tau, M_2 A_2 \bar\tau;   \frac{\rho_2^2}{|c_2 A_2 \tau + d_2|^2} \frac{\mu}{ n^2}  \Big)\notag \\
& = \sum_{A_2}\, Z \Big( A_2 \tau,  A_2 \bar\tau;   \frac{\rho_2^2}{n^2} \mu   \Big),\label{TnZA2}
}
where we used \eqref{theorem1} and \eqref{theorem2} in the second line, while in the third line we used the modular invariance of $Z(\tau, \bar\tau;\mu)$. Note that since the map between $A_1$ and $A_2$ in \eqref{theorem1} is one-to-one, the sum over $A_1$ in the first line of \eqref{TnZA2} can be written as a sum over $A_2$ in the second line, and both sums run over all inequivalent elements of $\Gamma(n)$. The third line of \eqref{TnZA2} is nothing but $(T'_n Z) (\tau, \bar \tau; \mu)$ as written in \eqref{heckettbar2}, so that the Hecke transform of the partition function \eqref{heckettbar2} is modular invariant for any positive integer $n$, namely
\eq{
(T'_n Z)\bigg(\frac{a \tau + b}{c \tau + d} , \frac{a \bar\tau + b}{c \bar\tau + d} ; \frac{\mu}{|c \tau + d|^2} \bigg) = (T'_n Z) (\tau, \bar \tau; \mu). \label{modularinvarianceHecke}
}

Let us now come back to the partition function of the symmetric orbifold $\symTT$. The latter can be made modular invariant by replacing each of the $Z(n\tau, n\bar\tau;\mu)$ terms in \eqref{untwistedZ} by their modular invariant completions by means of the Hecke operator, that is
\eq{
Z(n \tau, n\bar\tau; \mu)  \mapsto  (T'_n Z) ( \tau,  \bar \tau; \mu). 
} 
Consequently, the modular invariant partition function $Z_N(\tau, \bar\tau; \mu)$  of single-trace $T\bar T$-deformed CFTs is given by
\eq{
Z_N (\tau, \bar\tau; \mu) & =  \sum_{\{k_1, \dots, k_N\}} \frac{1}{\prod_{n = 1}^N n^{k_n} k_n ! } \prod_{n = 1}^N (T'_n Z)(\tau,  \bar \tau; \mu)^{k_n},\label{ZN}
}
where we used the fact that $(T'_1 Z) ( \tau,  \bar \tau; \mu) = Z ( \tau,  \bar \tau; \mu)$. 

The partition function \eqref{ZN} takes the same form as the partition function of a symmetric orbifold CFT \cite{Dijkgraaf:1996xw,Dijkgraaf:1998zd,Bantay:2000eq} except that the Hecke operator is generalized to act on the deformation parameter $\mu$ according to \eqref{heckettbar}. As described above, this is necessary to make the partition function modular invariant. We can use the generating functional of the cycle index of $S_N$ to write the generating functional $\mathcal Z(\tau,\bar\tau;\mu;p)$ of the partition function \eqref{ZN} such that
\eq{
\mathcal Z(\tau,\bar\tau;\mu;p) \coloneqq \sum_{N=0}^\infty p^N Z_N(\tau,\bar\tau;\mu) =  \exp \bigg( \sum_{n=1}^\infty \frac{p^n}{n} (T'_n Z)(\tau,\bar\tau;\mu)\bigg) ,\label{generatingfunctional}
}
where we have introduced the notation $Z_0 (\tau,\bar\tau;\mu) \equiv 1$ for convenience. Note that \eqref{generatingfunctional} also takes the same form as the generating functional of a symmetric orbifold CFT \cite{Dijkgraaf:1996xw,Bantay:2000eq} but with a generalized Hecke operator.

It is important to note that although \eqref{ZN} is modular invariant, it does not satisfy the differential equation \eqref{diffeqZ}. This is compatible with the fact that the spectrum of the symmetric orbifold includes sums of square roots, as well as twisted states, and is not equivalent to the spectrum of the $T\bar T$-deformed seed. Nevertheless, we note that each of the $(T'_n Z)(\tau,\bar\tau; \mu)$ terms in \eqref{ZN} satisfies the following differential equation
\eq{
n \p_{\mu} \big[(T_n'Z)(\tau, \bar \tau; \mu)\big] = \frac{1}{i\pi} \bigg[ (\tau - \bar \tau) \p_\tau \p_{\bar\tau} - \mu (\p_\tau - \p_{\bar\tau}) \p_{\mu} + \frac{2\mu}{\tau - \bar\tau} \p_{\mu} \bigg] (T_n'Z)(\tau, \bar \tau; \mu), \label{diffeqTnZ}
}
which resembles \eqref{diffeqZ} but with a rescaled deformation parameter $\mu \mapsto \mu/n$. In particular, the differential equation \eqref{diffeqTnZ} implies a differential equation for the generating functional \eqref{generatingfunctional}, which satisfies
\eq{
\!\!\! p\p_p \p_{\mu}\! \log  \mathcal Z(\tau,\bar\tau;\mu;p)  = \frac{1}{i\pi} \bigg[ (\tau - \bar \tau) \p_\tau \p_{\bar\tau} - \mu (\p_\tau - \p_{\bar\tau}) \p_{\mu} + \frac{2\mu}{\tau - \bar\tau} \p_{\mu} \bigg]\! \log \mathcal Z(\tau,\bar\tau;\mu;p). \label{diffeqgenZ}
}

A holographic derivation of the partition function has been previously carried out in~\cite{Hashimoto:2019hqo} by summing over the spectrum of winding strings on a linear dilaton background~\cite{Giveon:2017nie}. In contrast, our derivation of the partition function is purely field theoretical and does not rely on holography. Our result is also universal, in the sense that it applies to single-trace $T\bar T$ deformations of any symmetric orbifold CFT. Altogether, the fact that \eqref{generatingfunctional} matches the result obtained in \cite{Hashimoto:2019hqo} provides further evidence that the long string sector of string theory on a linear dilaton or TsT transformed background is holographically dual to a symmetric orbifold of a $T\bar T$-deformed CFT \cite{Giveon:2017nie,Apolo:2019zai}.


\subsection{The spectrum of twisted states}

We have seen that the partition function of the symmetric orbifold $\symTT$ \eqref{ZN} can be made modular invariant by means of the Hecke operator \eqref{heckettbar}. The latter introduces additional states --- the so-called twisted states --- to the symmetric orbifold whose spectrum we proceed to derive. 

Let us denote the single particle twisted states of $\symTT$ by $\phi_n$ ($n > 1$) where $n$ is the amount of twist.\footnote{A single particle state refers to a state that is not made of the product of states (other than the vacuum) while a multi particle state is made of the product of two or more single particle states.} In analogy with symmetric orbifold CFTs, we can associate each twisted state $\phi_n$ with a $\mathbb Z_n$ cycle of $S_N$, namely with the $(T'_n Z)(\tau,\bar\tau; \mu)$ factor in \eqref{ZN}. In order to see this, let us first expand the action of the Hecke operator on the partition function of the seed $T\bar T$-deformed CFT. For convenience, we denote the energies of the seed theory by $\cE_{L,R}(\mu)$ so that the seed partition function reads
\eq{
Z(\tau,\bar\tau;\mu) = \text{Tr}\Big( q^{\cE_L(\mu)} \bar{q}^{\cE_R( \mu)}\Big) = \sum_{\cE_L, \cE_R} \rho(\cE_L, \cE_R) e^{2\pi i \tau \cE_L(\mu) - 2\pi i \bar\tau \cE_R( \mu)}. \label{Zdef2}
}
On the other hand, the energies of $\symTT$ are denoted by $E_{L,R}(\mu)$. The action of the Hecke operator on $Z(\tau,\bar\tau;\mu)$ can then be expanded as follows
\eq{
\!\!\!\! \!\!\!\! \!\!\!\! (T'_n Z) ( \tau,  \bar \tau; \mu) & =  \sum_{\gamma | n} \sum_{ \cE_L, \cE_R}   \rho( \cE_L,  \cE_R)  q^{(n/\gamma^2)  \cE_L(\mu/\gamma^2)} \bar{q}^{(n/\gamma^2) \cE_R(\mu/\gamma^2)} \sum_{\a = 0}^{\gamma-1} e^{(2\pi i \alpha/\gamma)\J(0)} \notag \\
& =  \sum_{\gamma | n} \sum_{ \cE_L, \cE_R}   \rho( \cE_L,  \cE_R)  q^{(n/\gamma^2)  \cE_L(\mu/\gamma^2)} \bar{q}^{(n/\gamma^2)  \cE_R(\mu/\gamma^2)} \gamma \, \delta_{\J(0)}^{(\gamma)}, \label{heckeexp}
}
where in the first line we used the fact that the angular momentum of the seed theory $\J(\mu)$ is unchanged by the deformation, namely $ \cE_L(\mu/\gamma^2) -  \cE_R(\mu/\gamma^2) = \J(\mu/\gamma^2) = \J(0)$; while the $\delta_{\J(0)}^{(\gamma)} \coloneqq \delta_{\J(0)\,\textrm{mod}\,\gamma}$ term in the second line follows from the sum over roots of unity, which vanishes unless $\J(0)$ is a multiple of $\gamma$.

We now observe that when $n$ is a positive integer, the sum over the factors of $n$ in the Hecke transform \eqref{heckeexp} can be written as
\eqsp{
(T'_n Z) ( \tau,  \bar \tau; \mu) & = \sum_{ \cE_L, \cE_R}   \rho( \cE_L,  \cE_R)  q^{n  \cE_L(\mu)} \bar{q}^{n  \cE_R(\mu)} \\
&  \hspace{5pt} + \!\!\!\! \sum_{\substack{\gamma | n \\ \gamma \ne \{1, n\}} } \!\!  \sum_{ \cE_L, \cE_R}   \gamma \rho( \cE_L,  \cE_R)  q^{(n/\gamma^2)  \cE_L(\mu/\gamma^2)} \bar{q}^{(n/\gamma^2)  \cE_R(\mu/\gamma^2)}  \delta_{\J(0)}^{(\gamma)}\\
&  \hspace{5pt} +  \!\! \sum_{ \cE_L, \cE_R} n  \rho( \cE_L,  \cE_R)  q^{(1/n)  \cE_L(\mu/n^2)} \bar{q}^{(1/n)  \cE_R(\mu/n^2)} \delta_{\J(0)}^{(n)}. \label{heckeexp2}
}
The first line above is the contribution of a multi particle state from the untwisted sector that is made up of $n$ states from $n$ copies of the symmetric orbifold. The second line of \eqref{heckeexp2}, which is only present when $n$ is not a prime, corresponds to the contribution of multi particle states from the twisted sector that are made up of the product of $n/\gamma$ twisted states with twist $\gamma$ for each $\gamma \ne \{1,n\}$. Finally, the third line of \eqref{heckeexp2} corresponds to the sought-after contribution of the single particle twisted state $\phi_n$. This follows from the fact that the energy of this state cannot be expressed as the sum of energies of other states, in contrast to the multi particles states obtained from the first two lines.

The third line of \eqref{heckeexp2} implies that for each state in the $T\bar T$-deformed seed $\seedTT$ with energies $ \cE_{L,R}(\mu) \equiv E_{L,R}^{(1)}(\mu)$ satisfying $\J(0)\, \textrm{mod}\, n = 0$, there are $n$ single particle twisted states in $\symTT$ for each $n \in (1, N]$ whose deformed energies are given by
\eq{
 E_L^{(n)}(\mu) = \frac{1}{n}  E_L^{(1)}(\mu/n^2), \qquad  E_R^{(n)}(\mu) = \frac{1}{n}  E_R^{(1)}(\mu/n^2). \label{twistedstates}
} 
Up to the rescaling of the deformation parameter, the relation between the twisted states and untwisted states in $\symTT$ takes the same form as that in symmetric orbifold CFTs. This observation allows us to write the generating functional for the partition function of single-trace $T\bar T$-deformed CFTs in a similar way to the generating functional of symmetric orbifold CFTs \cite{Dijkgraaf:1996xw}, namely
\eq{
\mathcal Z(\tau, \bar\tau; \mu; p) =\prod_{n>0} \prod_{ \cE_{L}, \cE_R}\big (1-p^n q^{ \frac{1}{n}\cE_L(\mu/n^2)}\bar q^{ \frac{1}{n}\cE_R(\mu/n^2)} \big)^{-\rho( \cE_L, \cE_R) \delta_{\J(0)}^{(n)}},\label{generatingfunctional2}
}
where $\cE_{L,R}(\mu)$ and $\rho( \cE_L, \cE_R)$ are the energies and the density of states in the seed theory defined via \eqref{Zdef2}.

Using the relation \eqref{twistedstates} together with the expressions for the deformed energies \eqref{TTbarspectrum}, we can write the spectrum of twisted states in $\symTT$ in terms of the twisted states of the undeformed symmetric orbifold $\symCFT$ such that
\eqsp{
E_L^{(n)}(0) &= E_L^{(n)} (\mu) + \frac{2 \mu}{n} E_L^{(n)}(\mu)  E_R^{(n)}(\mu), \\
E_R^{(n)}(0) &= E_R^{(n)} (\mu) + \frac{2\mu}{n} E_L^{(n)}(\mu)  E_R^{(n)}(\mu). \label{spectrumZN}
}
The spectrum \eqref{spectrumZN} matches the spectrum of perturbative strings on TsT-transformed backgrounds where $n$ is identified with the winding of the string along the spatial circle \cite{Giveon:2017nie,Apolo:2019zai}. This is compatible with the fact that the partition function of $\symTT$ matches the partition function of long strings on a linear dilaton background, and provides further evidence for the correspondence.


\section{Single-trace $T\bar T$ partition functions at large $N$} \label{se:STpartitionfunction}

In this section we study the behavior of the partition function when the number of copies $N$ in the symmetric orbifold $\symTT$ is large, which corresponds to a large central charge in the undeformed symmetric orbifold $\symCFT$. More concretely, we will show that the partition function of single-trace $T\bar T$-deformed CFTs is universal in the large-$N$ limit without needing to impose any conditions on the density of light states. As a consequence of this, we will show that the density of light states saturates the sparseness condition, while the density of heavy states grows in a universal way analogous to the Cardy growth.

\subsection{Universality of the partition function}

Let us define the following version of the partition function of $\symTT$
\eq{
\til Z_N(\tau,\bar\tau;\mu) \coloneqq (q\bar q)^{-N \cE_{\vac}(\mu)/2}Z_N(\tau,\bar\tau;\mu), \label{ZNtilde}
}
where $\cE_{\vac}(\mu)$ denotes the vacuum energy of the deformed seed $\seedTT$. The total energy of the vacuum in $\symTT$ is therefore given by
\eq{
E_{\vac}(\mu) = N \cE_{\vac}(\mu) =  -\frac{N}{2\mu} \bigg(1 - \sqrt{1 - \frac{c_0 \mu}{3}}\,\bigg), \label{vacuumenergy2}
}
where $c_0$ denotes the central charge of the undeformed seed $\seedCFT$. In analogy with \cite{Hartman:2014oaa}, we will show that $\tilde Z_N (\tau, \bar\tau, \mu)$ is finite in the large-$N$ limit, which implies that $Z_N (\tau, \bar\tau, \mu)$ is dominated by the contribution of the vacuum without needing to make any assumptions on the sparseness of the light states.

In order to describe $\tilde Z_N(\tau, \bar\tau;\mu)$ in the large-$N$ limit, let us first rewrite the generating functional \eqref{generatingfunctional2} in the following way
\eq{
\!\!\!\! \mathcal Z (\tau, \bar\tau;\mu;p)  & = \prod_{n>0,  \cE_{L}, \cE_R}   \big (1- \til p^n q^{  \frac{1}{n} \cE_L(\mu/n^2) - \frac{n}{2} \cE_{\vac}(\mu) } \bar q^{ \, \frac{1}{n} \cE_R(\mu/n^2) -  \frac{n}{2} \cE_{\vac}(\mu) } \big)^{-\rho( \cE_L, \cE_R) \delta_{\J(0)}^{(n)}} \notag \\
& =  (1 - \til p)^{-1}  \sideset{}{'}\prod_{n>0,  \cE_{L}, \cE_R}  \big (1- \til p^n (q\bar q)^{-\frac{n}{2}\cE_{\vac}(\mu) } q^{ \frac{1}{n} \cE_L(\mu/n^2)} \bar q^{ \, \frac{1}{n}\cE_R(\mu/n^2)} \big)^{-\rho( \cE_L, \cE_R)\delta_{\J(0)}^{(n)}} \notag \\
& \equiv (1 - \til p)^{-1} R(\til p),
}
where $\til p = (q\bar q)^{ \frac{1}{2}\cE_{\vac}(\mu)} p$, the first factor in the second line comes from the contribution of the vacuum when $n = 1$, and the prime denotes the rest of the product which is denoted by  $R(\tilde p)$.  Let us formally expand $R(\tilde p)$ in a Taylor series such that $R(\tilde p)= \sum_{k=0}^\infty a_k\tilde p^k$ and $\mathcal Z(\tau, \bar\tau; \mu;p)$ can be written as
\eq{
\mathcal Z(\tau, \bar\tau; \mu;p) = \sum_{N=0}^\infty \tilde p^N\tilde Z_N=(1-\tilde p)^{-1}\sum_{k=0}^\infty a_k\tilde p^k.
}
Expanding the factor of $(1 - \tilde p)^{-1}$, it is not difficult to see that $\tilde Z_N(\tau,\bar\tau;\mu) = \sum_{k=0}^Na_k$ so that $\tilde Z_N(\tau,\bar\tau; \mu) =  R(1)$ in the limit $N \to \infty$. On the Lorentzian torus where $(\tau, \bar\tau) = (i \beta_L,-i\beta_R)$, this translates to
\eq{ 
\log  &  \tilde  Z_\infty  (\beta_L,\beta_R;\mu)  \notag \\
& = - \!\!\!\! \sideset{}{'}\sum_{n>0, \cE_L, \cE_R}  \!\! \rho( \cE_L,  \cE_R)\delta_{\J(0)}^{(n)} \log\Big( 1 - (q\bar q)^{  -  \frac{n}{2} \cE_{\vac}(\mu)  } q^{ \frac{1}{n}\cE_L(\mu/n^2)}\bar q^{\, \frac{1}{n}\cE_R(\mu/n^2) } \Big) \notag \\
&  = \sideset{}{'}\sum_{n>0, \cE_L, \cE_R}\sum_{k=1}^\infty \,\frac{1}{k} \rho( \cE_L,  \cE_R)\delta_{\J(0)}^{(n)} e^{2\pi k\big[\frac{n}{2}(\beta_L + \beta_R) \cE_{\vac}(\mu) -  \frac{1}{n} \beta_L \cE_L(\mu/n^2) -   \frac{1}{n} \beta_R\cE_R(\mu/n^2)\big]}, \label{logZNtilde}
}
where in the second line we expanded the logarithm using the fact that $\beta_{L,R} > 0$ and $\cE_{L,R}(\mu/n^2) > \frac{n^2}{2} \cE_{\vac}(\mu)$, the latter of which is true for all $n$ except the vacuum with $n = 1$, which is excluded from the sums above.

In order to show that $\tilde Z_\infty(\beta_L,\beta_R;\mu)$ is finite, we need to show that the sums in \eqref{logZNtilde} converge. We will assume that  $\beta_L \beta_R > 1$, since the proof in the case $\beta_L \beta_R < 1$ follows from modular invariance. Note that the sum in \eqref{logZNtilde} excludes the contribution from the vacuum when $n = 1$, so we treat this case separately. Omitting the factor of $\delta_{\J(0)}^{(n)}$ in \eqref{logZNtilde} then yields the following bound on the partition function
\eq{
\log   \tilde Z_\infty (\beta_L,\beta_R;\mu) < \sum_{k=1}^\infty \s_{1,k} + \sum_{n=2}^\infty \sum_{k=1}^\infty \s_{n,k}, \label{logZNtilde2}
}
where $\s_{1,k}$ and $\s_{n,k}$ are the $n = 1$ and $n \ge 2$ contributions to  \eqref{logZNtilde}, which are given by
\eq{
 \s_{n,k} \coloneqq\left\lbrace
\begin{aligned}
& \vphantom{\bigg(} \frac{1}{k} e^{\pi(\beta_L+\beta_R)k\cE_{\vac}(\mu)} Z'(k\beta_L,k\beta_R;\mu), \qquad \quad\,\,\,\, n=1,
	\\[1ex]
& \vphantom{\bigg(} \frac{1}{k}e^{\pi(\beta_L+\beta_R)nk \cE_{\vac}(\mu)}Z\Big(\frac{k\beta_L}{n},\frac{k\beta_R}{n};\frac{\mu}{n^2}\Big), \qquad n \ge 2,
\end{aligned}
 \right.\label{sigmandef}
}
with $Z'(\tau,\bar\tau;\mu)$ denoting the partition function of the seed theory without the vacuum. 

Let us first understand the behavior of $\sigma_{n,k}$ with $n \ge 2$. We begin by noting that the dominant contribution to the partition function $Z(\tau,\bar\tau;\mu)$ is given by $e^{-\pi (\beta_L + \beta_R) \cE_{\vac}(\mu)}$ when the inverse temperatures $\beta_{L,R}$ are both  large ($\beta_{L,R} \ll 1$), and by $e^{-\pi  (\beta'_L + \beta'_R) \cE_{\vac}(\mu')}$ when they are both small ($\beta_{L,R} \gg 1$) \cite{Datta:2018thy}. Assuming that both $\beta_{L,R}$ scale in the same way, the partition function $Z(\beta_L, \beta_R; \mu)$ of a $T\bar T$-deformed CFT can be bounded by
\eq{
Z(\beta_L, \beta_R; \mu)  & < \mathcal N e^{-\pi (\beta_L + \beta_R)   \cE_{\vac}(\mu)} e^{-\pi  (\beta'_L + \beta'_R)   \cE_{\vac}(\mu')}, \label{Zbound2}
}
where $\beta'_{L,R} = 1/\beta_{L,R}$, $\mu' = \mu/\beta_L\beta_R$, and $\mathcal N$ is a constant  that can be chosen to be large enough such that \eqref{Zbound2} holds for all temperatures. The bound on the partition function \eqref{Zbound2} implies that the $\s_{n,k}$ terms satisfy
\eq{
\s_{n, k} & <  \frac{1}{k} \mathcal N    e^{\pi n k(\beta_L + \beta_R)    \big( 1 - \frac{1}{n^2} - \frac{1}{k^2} \frac{1}{\beta_L \beta_R}\big) \cE_{\vac}(\mu)}, \qquad n \ge 2,
}
where we used the fact that $\cE_{\vac}\big(\frac{\mu'}{k^2}\big) >  \cE_{\vac}(\mu)$. The factor in the second parenthesis is positive for all $k \ge 2$ and  $\beta_L \beta_R > 1$, which together with the fact that the vacuum energy is negative, implies the following upper bound
\eq{
\s_{n, k} <    \frac{1}{k} \mathcal N  e^{\frac{\pi}{4} n k (\beta_L + \beta_R)   \big( 3 - \frac{1}{\beta_L \beta_R}\big)\cE_{\vac}(\mu) }, \qquad n \ge 2, \quad k \ge 2.\label{snk}
}
The right hand side of the above inequality can be summed over all $k\ge2$, the result of which is a series in $n$ which decreases exponentially as $n\to\infty$. As a result, the double sum in \eqref{logZNtilde2} converges for $n \ge 2$ and $k \ge 2$. 

While this approach is not useful when $k = 1$, it is possible to show that the sum $\sum_{n=2}^{\infty} \s_{n,1}$ converges in this case as well. Indeed, modular invariance of $Z \big(\frac{\beta_L}{n},\frac{\beta_R}{n};\frac{\mu}{n^2} \big)$ implies that for large $n$, the partition function can be approximated up to a constant $\widetilde{\mathcal N}$ by its large temperature limit  such that 
\eq{
Z\bigg(\frac{\beta_L}{n},\frac{\beta_R}{n};\frac{\mu}{n^2}\bigg) = Z \big( n \beta'_L, n \beta'_R;\mu' \big) < \widetilde{\mathcal N} e^{-\pi n(\beta'_L+\beta'_R) \cE_{\vac}(\mu')} \label{Zbound3}. 
}
Using $\cE_{\vac}(\mu')> \cE_{\vac}(\mu)$, we find that $\s_{n,1}$ is bounded by
\eq{
\s_{n,1}  < \widetilde{\mathcal N}  e^{\pi n (\beta_L + \beta_R)  \big(1  - \frac{1}{\beta_L \beta_R}\big)\cE_{\vac}(\mu)}, \qquad n \ge 2.
}
It follows that $\s_{n,1}$ decreases exponentially as $n\to \infty$ such that the sum over $n \ge 2$ with $k = 1$ also converges in this case.

Finally, let us consider the convergence of the single sum $\sum_{k=1}^{\infty} \s_{1,k}$ in \eqref{logZNtilde2}. Due to the exclusion of the vacuum, we can bound $Z'(\beta_L,\beta_R;\mu)$ as in \eqref{Zbound2} but with the vacuum replaced by the first excited state $\cE_1(\mu) > \cE_{\vac}(\mu)$ at low temperatures,
\eq{
Z'(\beta_L,\beta_R;\mu)<\mathcal N' e^{-\pi(\beta_L+\beta_R) \cE_1(\mu)}e^{-\pi(\beta'_L+\beta'_R)\cE_{\vac}(\mu')},
}
where $\mathcal N'$ is a constant. As a result, $\sigma_{1,k}$ is bounded by
\eq{
\s_{1,k} < \frac{1}{k}\mathcal N' e^{\pi k (\beta_L+\beta_R)\big(\cE_{\vac}(\mu)- \cE_1(\mu)-\frac{\cE_{\vac}(\mu')}{k^2\beta_L\beta_R}\big)}.
}
When $k$ is large enough, the $1/k^2$ term in the exponent can be dropped so that the exponent is negative and linear in $k$. Therefore, the sum over $k$ converges in this case as well.

We have shown that the partition function $\tilde Z_{N}(\tau, \bar\tau; \mu)$ is finite in the large-$N$ limit when $\beta_L \beta_R > 1$. This result can be extended to the high temperature region $\beta_L \beta_R < 1$ using the modular invariance of $Z(\tau, \bar\tau;\mu)$. Finally, using the relation between  $\tilde Z_{N}(\tau, \bar\tau; \mu)$ and $Z_{N}(\tau, \bar\tau; \mu)$ in \eqref{ZNtilde}, it follows that the partition function of single-trace $T\bar T$-deformed CFTs is universal in the large-$N$ limit such that
\eq{
 \log Z_N (\tau, \bar\tau; \mu) \approx \max \big\{\pi i (\tau - \bar\tau) E_{\vac}(\mu),  -\pi i \Big (\frac{1}{\tau} - \frac{1}{\bar\tau}\Big) E_{\vac}\Big(\frac{\mu}{\tau\bar\tau}\Big) \big\}, \qquad |\tau|^2 \ne 1.\label{approxZttbarST}
}
We see that this result is analogous to the one found in the double-trace case \eqref{approxZttbar}, except that here it is not necessary to assume sparseness of the light states. In fact, as described in the next section, \eqref{approxZttbarST} implies that the spectrum of $\textrm{Sym}^N \mathcal M_\mu$ is sparse, which can be shown to be a result of the orbifolding.


\subsection{Density of states}

We now show that the spectrum of single-trace $T\bar T$-deformed CFTs is universal in the large-$N$ limit. In particular, we show that the density of low energy states  saturates the sparseness condition \eqref{spa2-tt}, while the density of high energy states is given by a  universal formula analogous to Cardy.

Let us begin by noting that the universality of the partition function at large $N$ implies a universal bound on the density of states. In order to see this, let  $\rho_N(E_L, E_R)$ denote the density of states of $\symTT$ such that
\eq{
Z_N(\tau, \bar\tau; \mu) &= \sum_{ E_L,  E_R} \rho_N( E_L,  E_R) e^{-2\pi \beta_L   E_L(\mu)}e^{ -2\pi \beta_R E_R(\mu)}.
}
Using the fact that $\rho_N < Z_N (\tau, \bar\tau; \mu) e^{2\pi \beta_L   E_L(\mu)}e^{2\pi \beta_R  E_R(\mu)}$, it is not difficult to show that the density of states at large $N$ is bounded by
 \eq{
\log \rho_N( E_L,  E_R)& < \min_{\beta_L, \beta_R} \big\{ \log Z_N + 2 \pi \beta_L   E_L(\mu) + 2\pi \beta_R   E_R(\mu) \big\} \\
& \lesssim 4\pi \sqrt{\bigg(  E_L(\mu) - \frac{1}{2}  E_{\vac}(\mu)\bigg)\bigg( E_R(\mu) -  \frac{1}{2} E_{\vac}(\mu)\bigg)} \label{spa3},
}
where in the second line we have plugged in the large-$N$ approximation of the partition function \eqref{approxZttbarST} and optimized the bound. Note that the bound \eqref{spa3} applies to both high and low energy states. In particular, it reduces to the sparseness condition \eqref{spa2-tt} for the light states if we take $N = 1$ and $c_0 \to \infty$. However, the symmetric orbifold structure of the single-trace deformation allows us to derive this bound instead of assuming it. In the following, we will show that the bound \eqref{spa3} is compatible with the universal growth of single-trace $T\bar T$-deformed CFTs at high energies, and that it is furthermore saturated at low energies.

In analogy with the analysis of section \ref{se:entropy}, the asymptotic density of states can be obtained from the partition function \eqref{approxZttbarST} via the standard thermodynamic relations such that $\log \rho_N ( E_L,  E_R) =  S_N( E_L,  E_R)$ with the entropy $S_N( E_L,  E_R)$ given by
\eq{
S_N( E_L,  E_R) & \approx 2 \pi \bigg \{ \sqrt{\frac{c}{6}  E_L(\mu) \Big[ 1 + \frac{2\mu}{N}  E_R(\mu) \Big]} + \sqrt{\frac{c}{6}  E_R(\mu) \Big[ 1 + \frac{2\mu}{N}  E_L(\mu) \Big]}\, \bigg \}, \label{ttbarentropyST}
}
where $c = N c_0$ is the total central charge of the undeformed symmetric orbifold $\symCFT$. The formula \eqref{ttbarentropyST} is valid when $|\tau|^2<1$, which in the microcanonical ensemble corresponds to
\eq{
\frac{  E_L( \mu)  E_R( \mu)}{1 +  \frac{2\mu}{N} \big (  E_L( \mu)  +   E_R( \mu)\big )} >  \frac{  E_{\vac}(\mu) ^2 }{4\big(1 +  \frac{2\mu}{ N}  E_{\vac}(\mu) \big) }.  \label{cardyregimeST}
}
As a consistency check, we find that the entropy \eqref{ttbarentropyST} is indeed less than or equal to the bound \eqref{spa3} for the heavy states, with the inequality being saturated only when \eqref{cardyregimeST} is saturated. Note that the single-trace entropy formula \eqref{ttbarentropyST} and the range of validity \eqref{cardyregimeST} can be obtained from the double-trace versions \eqref{ttbarentropy2} and \eqref{cardyregime} by the replacement $c_0 \mapsto N c_0$ and $\mu \mapsto \mu/N$. This follows from the fact that the partition functions in the two theories, namely \eqref{approxZttbar} and \eqref{approxZttbarST}, are the same except for the vacuum energies,  the latter of which are related by the aforementioned rescaling as can be seen from \eqref{vacuumenergy} and \eqref{vacuumenergy2}.

For the states lying outside the region \eqref{cardyregimeST}, which includes the light states as well as the $T \bar T$ version of the enigmatic states, the bound \eqref{spa3} turns out to be actually saturated. In order to see this, we note that by expanding the generating functional \eqref{generatingfunctional} and comparing the coefficient of $p^N$, the density of states $\rho_{T'_n}(h_L, h_R)$ of the partition function $(T'_n Z)(\tau,\bar\tau; \mu) \coloneqq  (q\bar q)^{\frac{n}{2} \cE_{\vac}(\mu)}\sum_{h_L, h_R} \rho_{T'_n}( h_L,  h_R) q^{h_L(\mu)} \bar q^{h_R(\mu)}$ satisfies \cite{Hartman:2014oaa}
\eq{
\rho_N(h_L, h_R) \ge \sum_{n =1}^{N} \frac{1}{n} \rho_{T'_n}(h_L, h_R), \label{rhoTn}
} 
where $h_{L,R}(\mu)$ are the energies above the vacuum. Note that this relationship for the density of states only depends on the structure of the symmetric orbifold and does not depend on the details of the deformed seed $\seedTT$. Using the bound \eqref{spa3}, the relation \eqref{rhoTn} implies that the $n$th Hecke-transformed partition function $(T'_n Z)(\tau,\bar\tau; \mu)$ has a sparse light spectrum. At large $n\le N$, the fact that $(T'_n Z)(\tau,\bar\tau; \mu)$ is modular invariant allows us to go through the argument of section \ref{se:partitionfunction} and obtain a universal approximation to the partition function similar to \eqref{approxZttbar} and \eqref{approxZttbarST}, with the vacuum energy replaced by $n\cE_{\vac}(\mu)$. Consequently, the density of states of $(T'_n Z)(\tau,\bar\tau; \mu)$ at large $n$ is given by
\eq{
\log \rho_{T'_n}(h_L, h_R) \approx S_n \bigg(h_L + \frac{n}{2}  \cE_{\vac}, h_R + \frac{n}{2}  \cE_{\vac} \bigg). \label{Sn}
}
In terms of the $h_{L,R}(\mu)$ variables, the constraint \eqref{cardyregimeST} reads
\eq{
\frac{h_L(\mu) h_R(\mu)}{h_L(\mu) + h_R(\mu)} > -\frac{n  \cE_{\vac}(\mu) \big(1+\mu { \cE_{\vac}(\mu)}  \big)}{2\big(1 + 2\mu { \cE_{\vac}(\mu)}\big)}. \label{hLhRregion}
}

The maximum value of \eqref{Sn}  is $S_{n_*} (h_L, h_R)  = 4\pi \sqrt{h_L(\mu) h_R(\mu)}$, where $n_*\le N$ is reached when \eqref{hLhRregion} is saturated, namely
\eq{
n^* = - \frac{2\big(1 + 2\mu { \cE_{\vac}(\mu)}\big) h_L(\mu) h_R(\mu)}{\cE_{\vac}(\mu)  \big(1+\mu { \cE_{\vac}(\mu)}  \big)\big(h_L(\mu) + h_R(\mu)\big)}.\label{nstar}
}
 Hence, if we assume that the energies $h_{L,R}(\mu)$ are of $\mathcal O(N)$, then  \eqref{rhoTn} implies that the density of states $\rho_N(h_L, h_R)$ at large $N$ is bounded from below by
\eq{
\log \rho_N(h_L, h_R) \gtrsim \log \rho_{T'_{n_*}}(h_L, h_R) \approx 4\pi \sqrt{h_L(\mu) h_R (\mu)}.
}
In terms of the energies $E_{L,R}(\mu) = h_{L,R}(\mu) + \frac{1}{2} E_{\vac}(\mu)$, this statement translates to
\eq{
\log \rho_N(h_L, h_R) \gtrsim 4\pi \sqrt{\bigg(  E_L(\mu) - \frac{1}{2}  E_{\vac}(\mu)\bigg)\bigg( E_R(\mu) -  \frac{1}{2} E_{\vac}(\mu)\bigg)}. \label{lowerbound}
}
This expression is valid when $n^* \le N$, impliying that the energies $E_{L,R}(\mu)$ satisfy 
\eq{ \label{cardyregimeST2}
\frac{  E_L( \mu)  E_R( \mu)}{1 +  \frac{2\mu}{N} \big (  E_L( \mu)  +   E_R( \mu)\big )} \le \frac{ E_{\vac}(\mu) ^2 }{4\big(1 +  \frac{2\mu}{N}  E_{\vac}(\mu) \big) },
}
which lie in the complement of the high energy states \eqref{cardyregimeST}. 

We have shown that for energies in the region \eqref{cardyregimeST2}, the density of states at large $N$ is bounded from above and below by the same factor, namely by \eqref{spa3} and \eqref{lowerbound}. It follows that the density of low energy states satisfies
\eq{
\log \rho_N( E_L,  E_R) \approx 4\pi \sqrt{\bigg(  E_L(\mu) - \frac{1}{2}  E_{\vac}(\mu)\bigg)\bigg( E_R(\mu) -  \frac{1}{2} E_{\vac}(\mu)\bigg)}. \label{spa4}
}
This is the same expression found in the right hand side of the sparseness condition \eqref{spa2-tt}, meaning that the orbifolding procedure has made the low energy spectrum sparse. This observation is independent of the seed $\seedTT$, and was originally shown in the context of symmetric orbifold CFTs in \cite{Keller:2011xi,Hartman:2014oaa}. We conclude that the spectrum of single-trace $T\bar T$-deformed CFTs is universal in the large-$N$ limit such that the spectrum saturates the sparseness bound \eqref{spa4} when the energies lie in \eqref{cardyregimeST2}, or is otherwise given by the Cardy-like formula \eqref{ttbarentropyST}  when the energies lie in \eqref{cardyregimeST}.


 \bigskip

\section*{Acknowledgments}
We are grateful to Alejandra Castro, Hongjie Chen, Pengxiang Hao, Wenxin Lai, and Fengjun Xu for helpful discussions. LA thanks the Asia Pacific Center for Theoretical Physics (APCTP) for hospitality during the focus program ``Integrability, Duality and Related Topics", as well as the Korea Institute for Advanced Study (KIAS) for hospitality during the ``East Asia Joint Workshop on Fields and Strings 2022'', where part of this work was completed. BY thanks the Tsinghua Sanya International Mathematics Forum for hospitality during the workshop and research-in-team program on “Black holes, Quantum Chaos, and Solvable Quantum Systems”. The work of LA was supported by the Dutch Research Council (NWO) through the Scanning New Horizons programme (16SNH02). WS is supported by the national key research and development program of China No.~2020YFA0713000 and the Beijing Municipal Natural Science Foundation No.~Z180003. BY is supported by NSFC Grant No.~11735001.
 


\bigskip

\appendix


\ifprstyle
	\bibliographystyle{apsrev4-1}
\else
	\bibliographystyle{JHEP}
\fi

\bibliography{jtbar}

\providecommand{\href}[2]{#2}\begingroup\raggedright\begin{thebibliography}{10}

\bibitem{Strominger:1996sh}
A.~Strominger and C.~Vafa, \emph{{Microscopic origin of the Bekenstein-Hawking
  entropy}}, \href{https://doi.org/10.1016/0370-2693(96)00345-0}{\emph{Phys.
  Lett.} {\bfseries B379} (1996) 99}
  [\href{https://arxiv.org/abs/hep-th/9601029}{{\ttfamily hep-th/9601029}}].

\bibitem{Strominger:1997eq}
A.~Strominger, \emph{{Black hole entropy from near horizon microstates}},
  \href{https://doi.org/10.1088/1126-6708/1998/02/009}{\emph{JHEP} {\bfseries
  02} (1998) 009} [\href{https://arxiv.org/abs/hep-th/9712251}{{\ttfamily
  hep-th/9712251}}].

\bibitem{Cardy:1986ie}
J.~L. Cardy, \emph{{Operator Content of Two-Dimensional Conformally Invariant
  Theories}}, \href{https://doi.org/10.1016/0550-3213(86)90552-3}{\emph{Nucl.
  Phys.} {\bfseries B270} (1986) 186}.

\bibitem{Hartman:2014oaa}
T.~Hartman, C.~A. Keller and B.~Stoica, \emph{{Universal Spectrum of 2d
  Conformal Field Theory in the Large c Limit}},
  \href{https://doi.org/10.1007/JHEP09(2014)118}{\emph{JHEP} {\bfseries 09}
  (2014) 118} [\href{https://arxiv.org/abs/1405.5137}{{\ttfamily 1405.5137}}].

\bibitem{Benini:2015eyy}
F.~Benini, K.~Hristov and A.~Zaffaroni, \emph{{Black hole microstates in
  AdS$_{4}$ from supersymmetric localization}},
  \href{https://doi.org/10.1007/JHEP05(2016)054}{\emph{JHEP} {\bfseries 05}
  (2016) 054} [\href{https://arxiv.org/abs/1511.04085}{{\ttfamily
  1511.04085}}].

\bibitem{Cabo-Bizet:2018ehj}
A.~Cabo-Bizet, D.~Cassani, D.~Martelli and S.~Murthy, \emph{{Microscopic origin
  of the Bekenstein-Hawking entropy of supersymmetric AdS$_{5}$ black holes}},
  \href{https://doi.org/10.1007/JHEP10(2019)062}{\emph{JHEP} {\bfseries 10}
  (2019) 062} [\href{https://arxiv.org/abs/1810.11442}{{\ttfamily
  1810.11442}}].

\bibitem{Choi:2018hmj}
S.~Choi, J.~Kim, S.~Kim and J.~Nahmgoong, \emph{{Large AdS black holes from
  QFT}},  \href{https://arxiv.org/abs/1810.12067}{{\ttfamily 1810.12067}}.

\bibitem{Benini:2018ywd}
F.~Benini and P.~Milan, \emph{{Black holes in 4d $\mathcal{N}=4$
  Super-Yang-Mills}},  \href{https://arxiv.org/abs/1812.09613}{{\ttfamily
  1812.09613}}.

\bibitem{Guica:2008mu}
M.~Guica, T.~Hartman, W.~Song and A.~Strominger, \emph{{The Kerr/CFT
  Correspondence}},
  \href{https://doi.org/10.1103/PhysRevD.80.124008}{\emph{Phys. Rev.}
  {\bfseries D80} (2009) 124008}
  [\href{https://arxiv.org/abs/0809.4266}{{\ttfamily 0809.4266}}].

\bibitem{Smirnov:2016lqw}
F.~A. Smirnov and A.~B. Zamolodchikov, \emph{{On space of integrable quantum
  field theories}},
  \href{https://doi.org/10.1016/j.nuclphysb.2016.12.014}{\emph{Nucl. Phys.}
  {\bfseries B915} (2017) 363}
  [\href{https://arxiv.org/abs/1608.05499}{{\ttfamily 1608.05499}}].

\bibitem{Zamolodchikov:2004ce}
A.~B. Zamolodchikov, \emph{{Expectation value of composite field T anti-T in
  two-dimensional quantum field theory}},
  \href{https://arxiv.org/abs/hep-th/0401146}{{\ttfamily hep-th/0401146}}.

\bibitem{Cavaglia:2016oda}
A.~Cavagli\`a, S.~Negro, I.~M. Sz\'ecs\'enyi and R.~Tateo, \emph{{$T
  \bar{T}$-deformed 2D Quantum Field Theories}},
  \href{https://doi.org/10.1007/JHEP10(2016)112}{\emph{JHEP} {\bfseries 10}
  (2016) 112} [\href{https://arxiv.org/abs/1608.05534}{{\ttfamily
  1608.05534}}].

\bibitem{Guica:2017lia}
M.~Guica, \emph{{An integrable Lorentz-breaking deformation of two-dimensional
  CFTs}}, \href{https://doi.org/10.21468/SciPostPhys.5.5.048}{\emph{SciPost
  Phys.} {\bfseries 5} (2018) 048}
  [\href{https://arxiv.org/abs/1710.08415}{{\ttfamily 1710.08415}}].

\bibitem{LeFloch:2019rut}
B.~Le~Floch and M.~Mezei, \emph{{Solving a family of $T\bar{T}$-like
  theories}},  \href{https://arxiv.org/abs/1903.07606}{{\ttfamily 1903.07606}}.

\bibitem{Chakraborty:2019mdf}
S.~Chakraborty, A.~Giveon and D.~Kutasov, \emph{{$T\bar{T}$, $J\bar{T}$,
  $T\bar{J}$ and String Theory}},
  \href{https://doi.org/10.1088/1751-8121/ab3710}{\emph{J. Phys.} {\bfseries
  A52} (2019) 384003} [\href{https://arxiv.org/abs/1905.00051}{{\ttfamily
  1905.00051}}].

\bibitem{Frolov:2019xzi}
S.~Frolov, \emph{{$T{\overline T}$, $\widetilde JJ$, $JT$ and $\widetilde JT$
  deformations}},  \href{https://arxiv.org/abs/1907.12117}{{\ttfamily
  1907.12117}}.

\bibitem{Dubovsky:2012wk}
S.~Dubovsky, R.~Flauger and V.~Gorbenko, \emph{{Solving the Simplest Theory of
  Quantum Gravity}}, \href{https://doi.org/10.1007/JHEP09(2012)133}{\emph{JHEP}
  {\bfseries 09} (2012) 133} [\href{https://arxiv.org/abs/1205.6805}{{\ttfamily
  1205.6805}}].

\bibitem{Dubovsky:2013ira}
S.~Dubovsky, V.~Gorbenko and M.~Mirbabayi, \emph{{Natural Tuning: Towards A
  Proof of Concept}},
  \href{https://doi.org/10.1007/JHEP09(2013)045}{\emph{JHEP} {\bfseries 09}
  (2013) 045} [\href{https://arxiv.org/abs/1305.6939}{{\ttfamily 1305.6939}}].

\bibitem{Dubovsky:2017cnj}
S.~Dubovsky, V.~Gorbenko and M.~Mirbabayi, \emph{{Asymptotic fragility, near
  AdS$_{2}$ holography and $ T\overline{T} $}},
  \href{https://doi.org/10.1007/JHEP09(2017)136}{\emph{JHEP} {\bfseries 09}
  (2017) 136} [\href{https://arxiv.org/abs/1706.06604}{{\ttfamily
  1706.06604}}].

\bibitem{Datta:2018thy}
S.~Datta and Y.~Jiang, \emph{{$T\bar{T}$ deformed partition functions}},
  \href{https://doi.org/10.1007/JHEP08(2018)106}{\emph{JHEP} {\bfseries 08}
  (2018) 106} [\href{https://arxiv.org/abs/1806.07426}{{\ttfamily
  1806.07426}}].

\bibitem{Aharony:2018bad}
O.~Aharony, S.~Datta, A.~Giveon, Y.~Jiang and D.~Kutasov, \emph{{Modular
  invariance and uniqueness of $T\bar{T}$ deformed CFT}},
  \href{https://arxiv.org/abs/1808.02492}{{\ttfamily 1808.02492}}.

\bibitem{Cardy:2018sdv}
J.~Cardy, \emph{{The $ T\overline{T} $ deformation of quantum field theory as
  random geometry}}, \href{https://doi.org/10.1007/JHEP10(2018)186}{\emph{JHEP}
  {\bfseries 10} (2018) 186}
  [\href{https://arxiv.org/abs/1801.06895}{{\ttfamily 1801.06895}}].

\bibitem{Callebaut:2019omt}
N.~Callebaut, J.~Kruthoff and H.~Verlinde, \emph{{$ T\overline{T} $ deformed
  CFT as a non-critical string}},
  \href{https://doi.org/10.1007/JHEP04(2020)084}{\emph{JHEP} {\bfseries 04}
  (2020) 084} [\href{https://arxiv.org/abs/1910.13578}{{\ttfamily
  1910.13578}}].

\bibitem{Tolley:2019nmm}
A.~J. Tolley, \emph{{$T \bar T$ Deformations, Massive Gravity and Non-Critical
  Strings}},  \href{https://arxiv.org/abs/1911.06142}{{\ttfamily 1911.06142}}.

\bibitem{Giribet:2017imm}
G.~Giribet, \emph{{$T\bar{T}$-deformations, AdS/CFT and correlation
  functions}}, \href{https://doi.org/10.1007/JHEP02(2018)114}{\emph{JHEP}
  {\bfseries 02} (2018) 114}
  [\href{https://arxiv.org/abs/1711.02716}{{\ttfamily 1711.02716}}].

\bibitem{Donnelly:2018bef}
W.~Donnelly and V.~Shyam, \emph{{Entanglement entropy and $T \overline{T}$
  deformation}},
  \href{https://doi.org/10.1103/PhysRevLett.121.131602}{\emph{Phys. Rev. Lett.}
  {\bfseries 121} (2018) 131602}
  [\href{https://arxiv.org/abs/1806.07444}{{\ttfamily 1806.07444}}].

\bibitem{Chen:2018eqk}
B.~Chen, L.~Chen and P.-X. Hao, \emph{{Entanglement entropy in
  $T\overline{T}$-deformed CFT}},
  \href{https://doi.org/10.1103/PhysRevD.98.086025}{\emph{Phys. Rev.}
  {\bfseries D98} (2018) 086025}
  [\href{https://arxiv.org/abs/1807.08293}{{\ttfamily 1807.08293}}].

\bibitem{Jeong:2019ylz}
H.-S. Jeong, K.-Y. Kim and M.~Nishida, \emph{{Entanglement and R\'enyi entropy
  of multiple intervals in $T\overline{T}$-deformed CFT and holography}},
  \href{https://doi.org/10.1103/PhysRevD.100.106015}{\emph{Phys. Rev. D}
  {\bfseries 100} (2019) 106015}
  [\href{https://arxiv.org/abs/1906.03894}{{\ttfamily 1906.03894}}].

\bibitem{Cardy:2019qao}
J.~Cardy, \emph{{$T\bar T$ deformation of correlation functions}},
  \href{https://doi.org/10.1007/JHEP12(2019)160}{\emph{JHEP} {\bfseries 12}
  (2019) 160} [\href{https://arxiv.org/abs/1907.03394}{{\ttfamily
  1907.03394}}].

\bibitem{Hartman:2018tkw}
T.~Hartman, J.~Kruthoff, E.~Shaghoulian and A.~Tajdini, \emph{{Holography at
  finite cutoff with a $T^2$ deformation}},
  \href{https://doi.org/10.1007/JHEP03(2019)004}{\emph{JHEP} {\bfseries 03}
  (2019) 004} [\href{https://arxiv.org/abs/1807.11401}{{\ttfamily
  1807.11401}}].

\bibitem{Taylor:2018xcy}
M.~Taylor, \emph{{TT deformations in general dimensions}},
  \href{https://arxiv.org/abs/1805.10287}{{\ttfamily 1805.10287}}.

\bibitem{Gross:2019ach}
D.~J. Gross, J.~Kruthoff, A.~Rolph and E.~Shaghoulian, \emph{{$T\overline{T}$
  in AdS$_2$ and Quantum Mechanics}},
  \href{https://arxiv.org/abs/1907.04873}{{\ttfamily 1907.04873}}.

\bibitem{Giveon:2017nie}
A.~Giveon, N.~Itzhaki and D.~Kutasov, \emph{{$ \mathrm{T}\overline{\mathrm{T}}
  $ and LST}}, \href{https://doi.org/10.1007/JHEP07(2017)122}{\emph{JHEP}
  {\bfseries 07} (2017) 122}
  [\href{https://arxiv.org/abs/1701.05576}{{\ttfamily 1701.05576}}].

\bibitem{McGough:2016lol}
L.~McGough, M.~Mezei and H.~Verlinde, \emph{{Moving the CFT into the bulk with
  $T\overline{T} $}},
  \href{https://doi.org/10.1007/JHEP04(2018)010}{\emph{JHEP} {\bfseries 04}
  (2018) 010} [\href{https://arxiv.org/abs/1611.03470}{{\ttfamily
  1611.03470}}].

\bibitem{Guica:2019nzm}
M.~Guica and R.~Monten, \emph{{$T\bar T$ and the mirage of a bulk cutoff}},
  \href{https://arxiv.org/abs/1906.11251}{{\ttfamily 1906.11251}}.

\bibitem{Apolo:2019zai}
L.~Apolo, S.~Detournay and W.~Song, \emph{{TsT, $T\bar{T}$ and black strings}},
  \href{https://doi.org/10.1007/JHEP06(2020)109}{\emph{JHEP} {\bfseries 06}
  (2020) 109} [\href{https://arxiv.org/abs/1911.12359}{{\ttfamily
  1911.12359}}].

\bibitem{Apolo:2018qpq}
L.~Apolo and W.~Song, \emph{{Strings on warped AdS$_{3}$ via $
  \mathrm{T}\bar{\mathrm{J}} $ deformations}},
  \href{https://doi.org/10.1007/JHEP10(2018)165}{\emph{JHEP} {\bfseries 10}
  (2018) 165} [\href{https://arxiv.org/abs/1806.10127}{{\ttfamily
  1806.10127}}].

\bibitem{Araujo:2018rho}
T.~Araujo, E.~\'O~Colg\'ain, Y.~Sakatani, M.~M. Sheikh-Jabbari and
  H.~Yavartanoo, \emph{{Holographic integration of $T \bar{T}$ \& $J \bar{T}$
  via $O(d,d)$}}, \href{https://doi.org/10.1007/JHEP03(2019)168}{\emph{JHEP}
  {\bfseries 03} (2019) 168}
  [\href{https://arxiv.org/abs/1811.03050}{{\ttfamily 1811.03050}}].

\bibitem{Borsato:2018spz}
R.~Borsato and L.~Wulff, \emph{{Marginal deformations of WZW models and the
  classical Yang-Baxter equation}},
  \href{https://doi.org/10.1088/1751-8121/ab1b9c}{\emph{J. Phys.} {\bfseries
  A52} (2019) 225401} [\href{https://arxiv.org/abs/1812.07287}{{\ttfamily
  1812.07287}}].

\bibitem{Apolo:2021wcn}
L.~Apolo and W.~Song, \emph{{TsT, black holes, and $ T\overline{T} $ + $
  J\overline{T} $ + $ T\overline{J} $}},
  \href{https://doi.org/10.1007/JHEP04(2022)177}{\emph{JHEP} {\bfseries 04}
  (2022) 177} [\href{https://arxiv.org/abs/2111.02243}{{\ttfamily
  2111.02243}}].

\bibitem{Hashimoto:2019hqo}
A.~Hashimoto and D.~Kutasov, \emph{{Strings, Symmetric Products, $T \bar{T}$
  deformations and Hecke Operators}},
  \href{https://arxiv.org/abs/1909.11118}{{\ttfamily 1909.11118}}.

\bibitem{Chakraborty:2018vja}
S.~Chakraborty, A.~Giveon and D.~Kutasov, \emph{{$J\overline{T}$ deformed
  CFT$_{2}$ and string theory}},
  \href{https://doi.org/10.1007/JHEP10(2018)057}{\emph{JHEP} {\bfseries 10}
  (2018) 057} [\href{https://arxiv.org/abs/1806.09667}{{\ttfamily
  1806.09667}}].

\bibitem{Apolo:2019yfj}
L.~Apolo and W.~Song, \emph{{Heating up holography for single-trace $J\bar{T}$
  deformations}}, \href{https://doi.org/10.1007/JHEP01(2020)141}{\emph{JHEP}
  {\bfseries 01} (2020) 141}
  [\href{https://arxiv.org/abs/1907.03745}{{\ttfamily 1907.03745}}].

\bibitem{Aharony:2018vux}
O.~Aharony and T.~Vaknin, \emph{{The $T\bar{T}$ deformation at large central
  charge}}, \href{https://doi.org/10.1007/JHEP05(2018)166}{\emph{JHEP}
  {\bfseries 05} (2018) 166}
  [\href{https://arxiv.org/abs/1803.00100}{{\ttfamily 1803.00100}}].

\bibitem{Klemm:1990df}
A.~Klemm and M.~G. Schmidt, \emph{{Orbifolds by Cyclic Permutations of Tensor
  Product Conformal Field Theories}},
  \href{https://doi.org/10.1016/0370-2693(90)90164-2}{\emph{Phys. Lett. B}
  {\bfseries 245} (1990) 53}.

\bibitem{Dijkgraaf:1996xw}
R.~Dijkgraaf, G.~W. Moore, E.~P. Verlinde and H.~L. Verlinde, \emph{{Elliptic
  genera of symmetric products and second quantized strings}},
  \href{https://doi.org/10.1007/s002200050087}{\emph{Commun. Math. Phys.}
  {\bfseries 185} (1997) 197}
  [\href{https://arxiv.org/abs/hep-th/9608096}{{\ttfamily hep-th/9608096}}].

\bibitem{Dijkgraaf:1998zd}
R.~Dijkgraaf, \emph{{Fields, strings, matrices and symmetric products}},
  \href{https://arxiv.org/abs/hep-th/9912104}{{\ttfamily hep-th/9912104}}.

\bibitem{Bantay:2000eq}
P.~Bantay, \emph{{Symmetric products, permutation orbifolds and discrete
  torsion}}, \href{https://doi.org/10.1023/A:1024453119772}{\emph{Lett. Math.
  Phys.} {\bfseries 63} (2003) 209}
  [\href{https://arxiv.org/abs/hep-th/0004025}{{\ttfamily hep-th/0004025}}].

\bibitem{Apostol:1990}
T.~M. Apostol, \emph{{Modular Functions and Dirichlet Series in Number
  Theory}}. Springer-Verlag, New York, 1990.

\bibitem{Keller:2011xi}
C.~A. Keller, \emph{{Phase transitions in symmetric orbifold CFTs and
  universality}}, \href{https://doi.org/10.1007/JHEP03(2011)114}{\emph{JHEP}
  {\bfseries 03} (2011) 114} [\href{https://arxiv.org/abs/1101.4937}{{\ttfamily
  1101.4937}}].

\end{thebibliography}\endgroup



\end{document}
